\documentclass[twocolumn,showpacs,showkeys,preprintnumbers,amsmath,amssymb,aps,nofootinbib]{revtex4}
\usepackage{color}
\usepackage{indentfirst}
\usepackage{amsmath,amsthm,amssymb}
\usepackage{graphicx}
\usepackage[active]{srcltx}
\usepackage{hyperref}
\usepackage{tabularx}
\usepackage{setspace}

      \def\di{\displaystyle}
      
      \def\bS{{\bf S}}
      \def\bl{{\bf l}}
      \def\bp{{\bf p}}
      \def\bq{{\bf q}}
      \def\br{{\bf r}}

      \def\A{{\cal A}}

      \def\L{{\cal L}}
      
      \def\P{{\cal P}}
      \def\R{{\cal R}}

\begin{document}

\title{ A new type of nuclear collective motion -- the spin scissors mode}

\author{ E.B. Balbutsev\email{balbuts@theor.jinr.ru}, I.V. Molodtsova\email{molod@theor.jinr.ru}}
\affiliation{ Joint Institute for Nuclear Research, 141980 Dubna, Moscow Region,Russia }
\author{ P. Schuck}
\affiliation{Institut de Physique Nucl\'eaire, IN2P3-CNRS, Universit\'e Paris-Sud,
F-91406 Orsay C\'edex, France;\\
Laboratoire de Physique et Mod\'elisation des Milieux Condens\'es,
CNRS and Universit\'e Joseph Fourier,
25 avenue des Martyrs BP166, F-38042 Grenoble C\'edex 9, France  }

\begin{abstract}
The coupled dynamics of low lying modes and various giant resonances 
are studied with the help of the Wigner Function Moments method on 
the basis of Time Dependent Hartree-Fock equations in the harmonic 
oscillator model including spin-orbit potential plus 
quadrupole-quadrupole and spin-spin residual interactions. 
New low lying spin dependent modes are analyzed. Special attention is
paid to the spin scissors mode.
\end{abstract}

\pacs{ 21.10.Hw, 21.60.Ev, 21.60.Jz, 24.30.Cz } 
\keywords{spin; collective motion; scissors mode; giant resonances}

\maketitle

\section{Introduction}

The idea of the possible existence of the collective motion in deformed nuclei
similar to the scissors motion continues to attract the attention of physicists who extend it to various 
kinds of objects, not necessary nuclei, (for example, magnetic traps, see the review 
by Heyde at al~\cite{Heyd}) and invent new sorts of scissors, 
for example, the rotational oscillations of neutron skin against a proton-neutron core~\cite{Pena}.

The nuclear scissors mode was predicted \cite{Hilt}--\cite{Lo} 
as a counter-rotation of protons against neutrons in deformed nuclei.  
However, its collectivity turned out to be small. From RPA results which
 were in qualitative agreement with experiment, it was even questioned 
 whether this mode is collective at all \cite{Zaw,Sushkov}. 
Purely phenomenological models (such as, e.g., 
the two rotors model \cite{Lo2000}) and the sum rule approach~\cite{Lipp} 
did not clear up the situation in this 
respect. Finally in a very recent review \cite{Heyd} it is concluded 
that the scissors mode is "weakly collective, but strong 
on the single-particle scale" and further: "The weakly
collective scissors mode excitation has become an ideal test of models
-- especially microscopic models -- of nuclear vibrations. Most models
are usually calibrated to reproduce properties of strongly collective
excitations (e.g. of $J^{\pi}=2^+$ or $3^-$ states, giant resonances,
...). Weakly-collective phenomena, however, force the models to make
genuine predictions and the fact that the transitions in question are
strong on the single-particle scale makes it impossible to dismiss
failures as a mere detail, especially in the light of the overwhelming
experimental evidence for them in many nuclei \cite{Kneis,Richt}."

The Wigner Function
Moments (WFM) or phase space moments method turns out to be very 
useful in this
situation. On the one hand it is a purely microscopic method, because
it is based on the Time Dependent Hartree-Fock (TDHF) equation. On the
other hand the method works with average values (moments) of operators
which have a direct relation to the considered phenomenon and, thus, make a 
natural bridge with the macroscopic description. This 
makes it an ideal instrument to describe the basic characteristics 
(energies and excitation probabilities) of collective excitations such as,
in particular, the scissors mode. Our investigations have shown that 
already the minimal set of collective variables, i.e. phase space 
moments up to quadratic order,
is sufficient to reproduce the most important property of the
scissors mode: its inevitable coexistence with the IsoVector Giant
Quadrupole Resonance (IVGQR) implying a deformation of the Fermi surface.

Further developments of the Wigner Function Moments
method, namely, the switch from TDHF
to Time Dependent Hartree-Fock Bogoliubov (TDHFB) equations, i.e. taking 
into account pair correlations, allowed
us to improve considerably the quantitative description of the 
scissors mode \cite{Malov,Urban}: for rare earth nuclei the energies are 
reproduced with
$\sim 10\%$ accuracy and B(M1) values were reduced by about a factor of two  
with respect to their non superfluid values. 
However, they remain about two times too high with respect to experiment.
We have suspected, that the reason of this last discrepancy is hidden in the spin 
degrees of freedom, which were so far ignored by the WFM method. One cannot exclude,
that due to spin dependent interactions some part of the 
force of M1 transitions is shifted to the energy region of 5-10 MeV,
where a 1$^+$ resonance of spin nature is observed~\cite{Zaw}. 

  In a recent paper \cite{BaMo} the WFM method was 
applied for the first time to solve the TDHF equations including spin
dynamics.
As a first step, only the spin-orbit interaction was included in the
consideration, as the most important
one among all possible spin dependent interactions because it enters 
into the mean field. This allows one to understand the structure
of necessary modifications of the method avoiding  cumbersome 
calculations.
 The most remarkable result was the discovery of a new type
of nuclear collective motion: rotational oscillations of "spin-up"
nucleons with respect of "spin-down" nucleons (the spin scissors mode). 
It turns out that the experimentally 
observed group of peaks in the energy interval 2-4 MeV corresponds 
very likely to
two different types of motion: the conventional (orbital) scissors mode and this new kind 
of mode, i.e. the spin scissors mode.

Three
low lying excitations of a new nature were found: isovector and 
isoscalar spin scissors and the excitation generated by the relative 
motion of the orbital angular momentum and the spin of the nucleus (they 
can change their absolute values and directions keeping the total spin 
unchanged). 
In the frame of the same approach ten high lying excitations were also 
obtained: well known isoscalar and isovector Giant Quadrupole Resonances (GQR), 
two resonances of
a new nature describing isoscalar and isovector quadrupole vibrations
of "spin-up" nucleons with respect of "spin-down" nucleons, and six
resonances which can be interpreted as spin flip modes of various 
kinds and multipolarity.

The obtained results are very interesting, however, they are only 
intermediate in our investigation of M1 modes. Our finite goal 
is to get reasonable agreement with experimental data for the
conventional scissors mode, especially for its B(M1) factors
which remain about two times too strong.
We should keep in mind that only the standard spin-orbit potential was taken into account in 
the paper~\cite{BaMo},
spin dependent residual
interactions being completely neglected.

The aim of this work is to get a qualitative understanding of the influence of the spin-spin
force on the new states analyzed in~\cite{BaMo}, as, for instance, the spin scissors mode. 
As a matter of fact 
we will find 
 that the spin-spin interaction does not change the general
picture of the positions of excitations described in \cite{BaMo}. It pushes all levels
up proportionally to its strength without changing their order. The most interesting
result concerns the B(M1) values of both scissors modes 
-- the spin-spin interaction strongly redistributes M1 strength in the favour
of the spin scissors mode. This is a very promising fact, because it shows that
after taking into account in addition pairing \cite{Ring}
one may achieve  
agreement with experiment. 

One of the main points of the present work will, indeed, be that we will be 
able to give a tentative explanation of a recent experimental finding 
\cite{Siem} where the B(M1) values in $^{233}$Th of the two low lying 
magnetic states are inverted in strength in favor of the lowest, i.e.,
the spin scissors mode, when cranking up the spin-spin interaction. 
Indeed, the explanation with respect to a triaxial deformation given in \cite{Siem} 
yields a stronger B(M1) value for the higher lying state, contrary to 
observation, as remarked by the authors themselves.

The paper is organized as follows.
In Sec. 2 the TDHF equations for the 2x2 density matrix are
formulated and their Wigner transform is found.
In Sec. 3 the model Hamiltonian is analyzed and the mean field generated by the spin-spin interaction is found.
In Sec. 4 the collective variables are defined and the respective 
dynamical equations are derived.
In Sec. 5 the results of our calculations of energies, B(M1) and B(E2)
values are discussed.
Lastly, remarks and the outlook are given in the conclusion section. The mathematical details are concentrated 
in appendices \ref{AppA}, \ref{AppB}.

\section{Wigner transformation of TDHF equation with spin}

The TDHF equation in operator form reads~\cite{Ring}
\begin{equation}
i\hbar\dot{\hat\rho}=[\hat h,\hat\rho].
\label{tHF}
\end{equation}
Let us consider its matrix form in coordinate 
space keeping all spin indices:
\begin{widetext}
\begin{eqnarray}
i\hbar<\br,s|\dot{\hat\rho}|\br'',s''> =
\sum_{s'}\int\!d^3r'\!\left(
<\br,s|\hat h|\br',s'><\br',s'|\hat\rho|\br'',s''> 
-<\br,s|\hat\rho|\br',s'><\br',s'|\hat h|\br'',s''>\right).
\label{HFmatr}
\end{eqnarray}\end{widetext}
We do not specify the isospin indices in order 
to make the formulae more transparent. They
will be re-introduced at the end.

These equations will be solved by the method of phase space (or Wigner
function) moments. To this end we will rewrite the expression 
(\ref{HFmatr}) with the help of the
Wigner transformation \cite{Ring}.
To make the formulae more readable we will not
write out the coordinate dependence $(\br,\bp)$ of the functions. 
With the conventional notation 
$$\uparrow \, \mbox{for}\quad s=\frac{1}{2} \quad \mbox{and}
\quad\downarrow \, \mbox{for}\quad s=-\frac{1}{2}$$
the Wigner transform of (\ref{HFmatr}) can be written as
\begin{widetext}
\begin{eqnarray}
      i\hbar\dot f^{\uparrow\uparrow} &=&i\hbar\{h^{\uparrow\uparrow},f^{\uparrow\uparrow}\}
+h^{\uparrow\downarrow}f^{\downarrow\uparrow}-f^{\uparrow\downarrow}h^{\downarrow\uparrow}
+\frac{i\hbar}{2}\{h^{\uparrow\downarrow},f^{\downarrow\uparrow}\}
-\frac{i\hbar}{2}\{f^{\uparrow\downarrow},h^{\downarrow\uparrow}\}
-\frac{\hbar^2}{8}\{\{h^{\uparrow\downarrow},f^{\downarrow\uparrow}\}\}
+\frac{\hbar^2}{8}\{\{f^{\uparrow\downarrow},h^{\downarrow\uparrow}\}\}+...,
\nonumber\\
      i\hbar\dot f^{\uparrow\downarrow} &=&
f^{\uparrow\downarrow}(h^{\uparrow\uparrow}-h^{\downarrow\downarrow})
+\frac{i\hbar}{2}\{(h^{\uparrow\uparrow}+h^{\downarrow\downarrow}),f^{\uparrow\downarrow}\}
-\frac{\hbar^2}{8}\{\{(h^{\uparrow\uparrow}-h^{\downarrow\downarrow}),f^{\uparrow\downarrow}\}\}
\nonumber\\&&
-h^{\uparrow\downarrow}(f^{\uparrow\uparrow}-f^{\downarrow\downarrow})
+\frac{i\hbar}{2}\{h^{\uparrow\downarrow},(f^{\uparrow\uparrow}+f^{\downarrow\downarrow})\}
+\frac{\hbar^2}{8}\{\{h^{\uparrow\downarrow},(f^{\uparrow\uparrow}-f^{\downarrow\downarrow})\}\}+....
\label{WHF}
\end{eqnarray}
where the functions $h$, $f$ are the Wigner transforms of $\hat h$, 
$\hat\rho$ respectively, $\{f,g\}$ is the Poisson
bracket of the functions $f$ and $g$ and
$\{\{f,g\}\}$ is their double Poisson bracket;
the dots stand for terms proportional to higher powers of $\hbar$.
The remaining two equations are obtained by the obvious change of arrows
$\uparrow \leftrightarrow \downarrow$.

It is useful to rewrite the above equations in terms of functions 
$f^+=f^{\uparrow\uparrow}+f^{\downarrow\downarrow}$,
$f^-=f^{\uparrow\uparrow}-f^{\downarrow\downarrow}$.
By analogy with isoscalar $f^{\rm n}+f^{\rm p}$ and isovector  
$f^{\rm n}-f^{\rm p}$ functions one can name the functions $f^+$ and 
$f^-$ as spin-scalar and spin-vector ones, respectively.
We have:
\begin{eqnarray}
      &&i\hbar\dot f^{+} =\frac{i\hbar}{2}\{h^+,f^+\}+\frac{i\hbar}{2}\{h^-,f^-\}
+i\hbar\{h^{\uparrow\downarrow},f^{\downarrow\uparrow}\}
+i\hbar\{h^{\downarrow\uparrow},f^{\uparrow\downarrow}\}+...,
\nonumber\\
      &&i\hbar\dot f^{-} =
\frac{i\hbar}{2}\{h^+,f^-\}+\frac{i\hbar}{2}\{h^-,f^+\}
-2h^{\downarrow\uparrow}f^{\uparrow\downarrow}+2h^{\uparrow\downarrow}f^{\downarrow\uparrow}
+\frac{\hbar^2}{4}\{\{h^{\downarrow\uparrow},f^{\uparrow\downarrow}\}\}
-\frac{\hbar^2}{4}\{\{h^{\uparrow\downarrow},f^{\downarrow\uparrow}\}\}+...,
\nonumber\\
      &&i\hbar\dot f^{\uparrow\downarrow} =
-h^{\uparrow\downarrow}f^-+h^-f^{\uparrow\downarrow}
+\frac{i\hbar}{2}\{h^{\uparrow\downarrow},f^+\}
+\frac{i\hbar}{2}\{h^+,f^{\uparrow\downarrow}\}
+\frac{\hbar^2}{8}\{\{h^{\uparrow\downarrow},f^-\}\}
-\frac{\hbar^2}{8}\{\{h^-,f^{\uparrow\downarrow}\}\}+...,
\nonumber\\
      &&i\hbar \dot f^{\downarrow\uparrow} =
h^{\downarrow\uparrow}f^- -h^-f^{\downarrow\uparrow}
+\frac{i\hbar}{2}\{h^{\downarrow\uparrow},f^+\}
+\frac{i\hbar}{2}\{h^+,f^{\downarrow\uparrow}\}
-\frac{\hbar^2}{8}\{\{h^{\downarrow\uparrow},f^-\}\}
+\frac{\hbar^2}{8}\{\{h^-,f^{\downarrow\uparrow}\}\}+...,
\label{WHF+}
\end{eqnarray}
where $h^{\pm}=h^{\uparrow\uparrow}\pm h^{\downarrow\downarrow}$.

\section{Model Hamiltonian}

 The microscopic Hamiltonian of the model, harmonic oscillator with 
spin-orbit potential plus separable quadrupole-quadrupole and 
spin-spin residual interactions is given by
\begin{eqnarray}
\label{Ham}
 H=\sum\limits_{i=1}^A\left[\frac{\hat\bp_i^2}{2m}+\frac{1}{2}m\omega^2\br_i^2
-\eta\hat \bl_i\hat \bS_i\right]+H_{qq}+H_{ss}
\end{eqnarray}
with
\begin{eqnarray}
\label{Hqq}
 &&H_{qq}=\!
\sum_{\mu=-2}^{2}(-1)^{\mu}
\left\{\bar{\kappa}
 \sum\limits_i^Z\!\sum\limits_j^N
q_{2-\mu}(\br_i)q_{2\mu}(\br_j)
+\frac{1}{2}\kappa
\left[\sum\limits_{i\neq j}^{Z}
 q_{2-\mu}(\br_i)q_{2\mu}(\br_j)
+\sum\limits_{i\neq j}^{N}
 q_{2-\mu}(\br_i)q_{2\mu}(\br_j)\right]\!\right\},\\
\label{Hss}
&&H_{ss}=\!\!
\sum_{\mu=-1}^{1}(-1)^{\mu}
\left\{\bar{\chi}
 \sum\limits_i^Z\!\sum\limits_j^N
\hat S_{-\mu}(i)\hat S_{\mu}(j)
+\frac{1}{2}\chi
\left[\sum\limits_{i\neq j}^{Z}
\hat S_{-\mu}(i)\hat S_{\mu}(j)
+\sum\limits_{i\neq j}^{N}
\hat S_{-\mu}(i)\hat S_{\mu}(j)\right]\!\right\}
\!\delta(\br_i-\br_j),
\end{eqnarray}
where $N$ and $Z$ are the numbers of neutrons and protons
and $\hat S_{\mu}$ are spin matrices \cite{Var}:
\begin{equation}
\hat S_1=-\frac{\hbar}{\sqrt2}{0\quad 1\choose 0\quad 0},\quad
\hat S_0=\frac{\hbar}{2}{1\quad\, 0\choose 0\, -\!1},\quad
\hat S_{-1}=\frac{\hbar}{\sqrt2}{0\quad 0\choose 1\quad 0}.
\label{S}
\end{equation}
The quadrupole operator $q_{2\mu}=\sqrt{16\pi/5}\,r^2Y_{2\mu}(\theta,\phi)$ 
can be written as the tensor product:
$q_{2\mu}(\br)=\sqrt6\{r\otimes r\}_{2\mu},$
where
$$
\{r\otimes r\}_{\lambda\mu}=\sum_{\sigma,\nu}
C_{1\sigma,1\nu}^{\lambda\mu}r_{\sigma}r_{\nu},$$

$r_{-1}, r_0, r_1$ are cyclic coordinates \cite{Var} and 
$C_{1\sigma,1\nu}^{\lambda\mu}$ is a Clebsch-Gordan
coefficient.

\subsection{Mean Field}

Let us analyze the mean field generated by this Hamiltonian.

\subsubsection{Spin-orbit Potential}

Written in cyclic coordinates, the spin-orbit part of the
Hamiltonian reads
$$\hat h_{ls}=-\eta\sum_{\mu=-1}^1(-)^{\mu}\hat l_{\mu}\hat S_{-\mu}
=-\eta{\quad\hat l_0\frac{\hbar}{2}\quad\; \hat l_{-1}\frac{\hbar}{\sqrt2} \choose 
 -\hat l_{1}\frac{\hbar}{\sqrt2}\; -\hat l_0\frac{\hbar}{2}},
$$
where \cite{Var}
\begin{equation}
\label{lqu}
\hat l_{\mu}=-\hbar\sqrt2\sum_{\nu,\alpha}C_{1\nu,1\alpha}^{1\mu}r_{\nu}\nabla_{\alpha}
\end{equation}
and
\begin{eqnarray}
&&\hat l_1=\hbar(r_0\nabla_1-r_1\nabla_0)=
-\frac{1}{\sqrt2}(\hat l_x+i\hat l_y),\quad
\hat l_0=\hbar(r_{-1}\nabla_1-r_1\nabla_{-1})=\hat l_z,
\nonumber\\
&&\hat l_{-1}=\hbar(r_{-1}\nabla_0-r_0\nabla_{-1})=
\frac{1}{\sqrt2}(\hat l_x-i\hat l_y),
\nonumber\\
&&\hat l_x=-i\hbar(y\nabla_z-z\nabla_y),\quad
\hat l_y=-i\hbar(z\nabla_x-x\nabla_z),\quad
\hat l_z=-i\hbar(x\nabla_y-y\nabla_x).
\label{lxyz}
\end{eqnarray}
 Matrix elements of $\hat h_{ls}$ in coordinate space can be obviously written as
\begin{eqnarray}
<\br_1,s_1|\hat h_{ls}|\br_2,s_2>
&=&-\frac{\hbar}{2}\eta(\br_1)\{\hat l_{0}(\br_1)[\delta_{s_1\uparrow}\delta_{s_2\uparrow}
-\delta_{s_1\downarrow}\delta_{s_2\downarrow}]
\nonumber\\
&&+\sqrt2\, \hat l_{-1}(\br_1)\delta_{s_1\uparrow}\delta_{s_2\downarrow}
-\sqrt2\, \hat l_{1}(\br_1)\delta_{s_1\downarrow}\delta_{s_2\uparrow}\}\delta(\br_1-\br_2).
\label{Hrr'}
\end{eqnarray}
The Wigner transform of (\ref{Hrr'}) reads \cite{BaMo}:
\begin{eqnarray}
 h_{ls}^{s_1s_2}(\br,\bp)
=-\frac{\hbar}{2}\eta\{l_{0}(\br,\bp)[\delta_{s_1\uparrow}\delta_{s_2\uparrow}
-\delta_{s_1\downarrow}\delta_{s_2\downarrow}]
+\sqrt2 l_{-1}(\br,\bp)\delta_{s_1\uparrow}\delta_{s_2\downarrow}
-\sqrt2 l_{1}(\br,\bp)\delta_{s_1\downarrow}\delta_{s_2\uparrow}\},
\label{Hrp}
\end{eqnarray}
where
$l_{\mu}=-i\sqrt2\sum_{\nu,\alpha}C_{1\nu,1\alpha}^{1\mu}r_{\nu}p_{\alpha}$.

\subsubsection{q-q interaction}

 The contribution of $H_{qq}$ to the mean field potential is easily
found by replacing one of the $q_{2\mu}$ operators by the average value.
We have
\begin{equation}
\label{potenirr}
V^{\tau}_{qq}=6\sum_{\mu}(-1)^{\mu}Z_{2-\mu}^{\tau +}\{r\otimes r\}_{2\mu}.
\end{equation}
 Here
\begin{equation}
\label{Z2mu}
Z_{2\mu}^{n+}=\kappa R_{2\mu}^{n+}
+\bar{\kappa}R_{2\mu}^{p+}\,,\quad
Z_{2\mu}^{p+}=\kappa R_{2\mu}^{p+}
+\bar{\kappa}R_{2\mu}^{n+},\quad  R_{\lambda\mu}^{\tau+}(t)=
\int d(\bp,\br)
\{r\otimes r\}_{\lambda\mu}f^{\tau+}(\br,\bp,t)
\end{equation}
 with
 $\int\! d(\bp,\br)\equiv
(2\pi\hbar)^{-3}\int\! d^3p\,\int\! d^3r$ and $\tau$ being the isospin index.

\subsubsection{Spin-spin interaction}

The analogous expression for $H_{ss}$ is found in the
standard way, with the Hartree-Fock contribution given \cite{Ring} by:
\begin{equation}
\label{MeFi}
\Gamma_{kk'}(t)=\sum_{ll'}\bar v_{kl'k'l}\rho_{ll'}(t),
\end{equation}
where $\bar v_{kl'k'l}$ is the antisymmetrized matrix element of the two
body interaction $v(1,2)$. Identifying the indices $k,k',l,l'$
with the set of coordinates $(\br,s,\tau)$, i.e. (position, spin, isospin),
one rewrites (\ref{MeFi}) as
\begin{eqnarray}
V^{HF}(\br_1,s_1,\tau_1;\br_1',s_1',\tau_1';t)
=\int\!d^3r_2\int\!d^3r_2'\sum_{s_2,s_2'}\sum_{\tau_2,\tau_2'}\hspace{7cm}
\nonumber\\
<\br_1,s_1,\tau_1;\br_2,s_2,\tau_2|\hat v|\br_1',s_1',\tau_1';\br_2',s_2',\tau_2'>_{a.s.}
\rho(\br_2',s_2',\tau_2';\br_2,s_2,\tau_2;t).
\nonumber
\end{eqnarray}
Let us consider the neutron-proton part of the spin-spin interaction. 
In this case
$$\hat v=v(\hat\br_1-\hat\br_2)
\sum_{\mu=-1}^{1}(-1)^{\mu}
\hat S_{-\mu}(1)\hat S_{\mu}(2)
\delta_{\tau_1p}\delta_{\tau_2n},$$
where $\hat\br_1$ is the position operator: $\hat\br_1|\br_1>=\br_1|\br_1>, \quad
<\br_1|\hat\br_1|\br_1'>=<\br_1|\br_1'>\br_1'=\delta(\br_1-\br_1')\br_1'.$

For the Hartree term one finds:
\begin{eqnarray}
<\br_1,s_1,\tau_1;\br_2,s_2,\tau_2|\hat v|\br_1',s_1',\tau_1';\br_2',s_2',\tau_2'>=
\delta(\br_1-\br_1')\delta(\br_2-\br_2')v(\br_1'-\br_2')
\nonumber\\
\sum_{\mu=-1}^{1}(-1)^{\mu}<s_1,\tau_1;s_2,\tau_2|
\hat S_{-\mu}(1)\hat S_{\mu}(2)
\delta_{\tau_1p}\delta_{\tau_2n}|s_1',\tau_1';s_2',\tau_2'>,
\nonumber
\end{eqnarray}
\begin{eqnarray}
V^{H}(\br_1,s_1,\tau_1;\br_1',s_1',\tau_1';t)
=\int\!d^3r_2\int\!d^3r_2'\sum_{s_2,s_2'}\sum_{\tau_2,\tau_2'}\hspace{7cm}
\nonumber\\
<\br_1,s_1,\tau_1;\br_2,s_2,\tau_2|\hat v|\br_1',s_1',\tau_1';\br_2',s_2',\tau_2'>
\rho(\br_2',s_2',\tau_2';\br_2,s_2,\tau_2;t)
\nonumber\\
=\delta_{\tau_1p}\delta_{\tau_1'p}\sum_{s_2,s_2'}\sum_{\mu=-1}^{1}(-1)^{\mu}
<s_1|\hat S_{-\mu}(1)|s_1'><s_2|\hat S_{\mu}(2)|s_2'>
\nonumber\\
\delta(\br_1-\br_1')\int\!d^3r_2v(\br_1-\br_2)
\rho(\br_2,s_2',n;\br_2,s_2,n;t).
\nonumber
\end{eqnarray}
The Fock term reads:
\begin{eqnarray}
<\br_1,s_1,\tau_1;\br_2,s_2,\tau_2|\hat v|\br_2',s_2',\tau_2';\br_1',s_1',\tau_1'>=
\delta(\br_1-\br_2')\delta(\br_2-\br_1')v(\br_2'-\br_1')
\nonumber\\
\sum_{\mu=-1}^{1}(-1)^{\mu}<s_1,\tau_1;s_2,\tau_2|
\hat S_{-\mu}(1)\hat S_{\mu}(2)
\delta_{\tau_1p}\delta_{\tau_2n}|s_2',\tau_2';s_1',\tau_1'>,
\nonumber
\end{eqnarray}
\begin{eqnarray}
V^{F}(\br_1,s_1,\tau_1;\br_1',s_1',\tau_1';t)
=-\int\!d^3r_2\int\!d^3r_2'\sum_{s_2,s_2'}\sum_{\tau_2,\tau_2'}\hspace{7cm}
\nonumber\\
<\br_1,s_1,\tau_1;\br_2,s_2,\tau_2|\hat v|\br_2',s_2',\tau_2';\br_1',s_1',\tau_1'>
\rho(\br_2',s_2',\tau_2';\br_2,s_2,\tau_2;t)
\nonumber\\
=-\delta_{\tau_1p}\delta_{\tau_1'n}\sum_{s_2,s_2'}
\sum_{\mu=-1}^{1}(-1)^{\mu}<s_1|\hat S_{-\mu}(1)|s_2'><s_2|\hat S_{\mu}(2)|s_1'>
\nonumber\\
v(\br_1-\br_1')\rho(\br_1,s_2',p;\br_1',s_2,n;t).
\nonumber
\end{eqnarray}
Taking into account the relations
$$
<s|\hat S_{-1}|s'>=\frac{\hbar}{\sqrt2}\delta_{s\downarrow}\delta_{s'\uparrow},
\qquad 
<s|\hat S_{0}|s'>=\frac{\hbar}{2}\delta_{s,s'}(\delta_{s\uparrow}-\delta_{s\downarrow}),
\qquad  
<s|\hat S_{1}|s'>=-\frac{\hbar}{\sqrt2}\delta_{s\uparrow}\delta_{s'\downarrow}
$$
 and $v(\br-\br')=\bar\chi\delta(\br-\br')$
one finds for the mean field generated by the proton-neutron 
part of~$H_{ss}$:
\begin{eqnarray}
\label{Gam}
\Gamma_{pn}(\br,s,\tau;\br',s',\tau';t)=\bar\chi\frac{\hbar^2}{4}
\Bigg\{
\delta_{\tau p}\delta_{\tau'p}\Big[
\delta_{s\downarrow}\delta_{s'\uparrow}\rho(\br,\downarrow,n;\br',\uparrow,n;t)
+\delta_{s\uparrow}\delta_{s'\downarrow}\rho(\br,\uparrow,n;\br',\downarrow,n;t)\Big]
\nonumber\\
-\delta_{\tau p}\delta_{\tau'n}\Big[
\delta_{s\downarrow}\delta_{s'\downarrow}\rho(\br,\uparrow,p;\br',\uparrow,n;t)+
\delta_{s\uparrow}\delta_{s'\uparrow}\rho(\br,\downarrow,p;\br',\downarrow,n;t)\Big]
\nonumber\\
+\frac{1}{2}\delta_{\tau p}\delta_{\tau'p}\left(\delta_{s\uparrow}\delta_{s'\uparrow}-
\delta_{s\downarrow}\delta_{s'\downarrow}\right)\Big[\rho(\br,\uparrow,n;\br',\uparrow,n;t)
-\rho(\br,\downarrow,n;\br',\downarrow,n;t)\Big]
\nonumber\\
+\frac{1}{2}\delta_{\tau p}\delta_{\tau'n}\Big[
\delta_{s\uparrow}\delta_{s'\downarrow}\rho(\br,\uparrow,p;\br',\downarrow,n;t)
+\delta_{s\downarrow}\delta_{s'\uparrow}\rho(\br,\downarrow,p;\br',\uparrow,n;t)
\nonumber\\
-\delta_{s\uparrow}\delta_{s'\uparrow}\rho(\br,\uparrow,p;\br',\uparrow,n;t)
-\delta_{s\downarrow}\delta_{s'\downarrow}\rho(\br,\downarrow,p;\br',\downarrow,n;t)\Big]
\Bigg\}\delta(\br-\br')
+\bar\chi\frac{\hbar^2}{4}\Bigg\{p\leftrightarrow n\Bigg\}\delta(\br-\br').
\end{eqnarray}
The expression for the mean field $\Gamma_{pp}(\br,s,\tau;\br',s',\tau';t)$ 
generated by the proton-proton part of $H_{ss}$ can be obtained from 
(\ref{Gam}) by replacing index $n$ by $p$  and the strength constant 
$\bar\chi$ by $\chi$. The proton mean field is defined as the sum 
of these two terms with $\tau=\tau'=p$. Its Wigner transform can be
written as
\begin{eqnarray}
\label{Vp}
V_{p}^{s s'}(\br,t)&=&
3\chi\frac{\hbar^2}{8}
\left\{
\delta_{s\downarrow}\delta_{s'\uparrow}n_p^{\downarrow\uparrow}+
\delta_{s\uparrow}\delta_{s'\downarrow}n_p^{\uparrow\downarrow}
-\delta_{s\downarrow}\delta_{s'\downarrow}n_p^{\uparrow\uparrow}
-\delta_{s\uparrow}\delta_{s'\uparrow}n_p^{\downarrow\downarrow}
\right\}
\nonumber\\
&+&\bar\chi\frac{\hbar^2}{8}
\left\{
2\delta_{s\downarrow}\delta_{s'\uparrow}n_n^{\downarrow\uparrow}+
2\delta_{s\uparrow}\delta_{s'\downarrow}n_n^{\uparrow\downarrow}
+(\delta_{s\uparrow}\delta_{s'\uparrow}-
\delta_{s\downarrow}\delta_{s'\downarrow})(n_n^{\uparrow\uparrow}-
n_n^{\downarrow\downarrow})
\right\},
\end{eqnarray}
where 
${\di n_{\tau}^{ss'}(\br,t)=\int\frac{d^3p}{(2\pi\hbar)^3}f^{ss'}_{\tau}(\br,\bp,t)}$.
The Wigner transform of the neutron mean field $V_n^{ss'}$ is 
obtained from (\ref{Vp}) by the obvious change of indices $p\leftrightarrow n$.
The Wigner function $f$ and density matrix 
$\rho$ are connected by the relation 
${\di f^{ss'}_{\tau\tau'}(\br,\bp,t)=\int\!d^3\! q\, e^{-i\bp\bq/\hbar}
\rho(\br_1,s,\tau;\br_2,s',\tau';t)}$, with $\bq=\br_1-\br_2$ and 
$\br=\frac{1}{2}(\br_1+\br_2)$. Integrating this relation over $\bp$
with $\tau'=\tau$ one finds:
$$n_{\tau}^{ss'}(\br,t)=\rho(\br,s,\tau;\br,s',\tau;t).$$
By definition the diagonal elements of the density matrix describe 
the proper densities. Therefore $n_{\tau}^{ss}(\br,t)$ is the density
of spin-up nucleons (if $s=\uparrow$) or spin-down nucleons 
(if $s=\downarrow$). Off diagonal in spin elements of the
density matrix $n_{\tau}^{ss'}(\br,t)$ are spin-flip characteristics
and can be called spin-flip densities.

\section{Equations of motion}

 Integrating the set of equations (\ref{WHF+}) over phase space 
with the weights 
\begin{equation}
W =\{r\otimes p\}_{\lambda\mu},\,\{r\otimes r\}_{\lambda\mu},\,
\{p\otimes p\}_{\lambda\mu}, \mbox{ and } 1
\label{weightfunctions}
\end{equation}
one gets dynamic equations for 
the following collective variables:
\begin{eqnarray}
&&L^{\tau\varsigma}_{\lambda\mu}(t)=\int\! d(\bp,\br) \{r\otimes p\}_{\lambda\mu}
f^{\tau\varsigma}(\br,\bp,t),\quad
R^{\tau\varsigma}_{\lambda\mu}(t)=\int\! d(\bp,\br) \{r\otimes r\}_{\lambda\mu}
f^{\tau\varsigma}(\br,\bp,t),\quad
\nonumber\\
&&P^{\tau\varsigma}_{\lambda\mu}(t)=\int\! d(\bp,\br) \{p\otimes p\}_{\lambda\mu}
f^{\tau\varsigma}(\br,\bp,t),\quad
F^{\tau\varsigma}(t)=\int\! d(\bp,\br)
f^{\tau\varsigma}(\br,\bp,t),\quad
\label{Varis}
\end{eqnarray}
 where 
$\varsigma\!=+,\,-,\,\uparrow\downarrow,\,\downarrow\uparrow.$
We already called the functions 
$f^+=f^{\uparrow\uparrow}+f^{\downarrow\downarrow}$ and
$f^-=f^{\uparrow\uparrow}-f^{\downarrow\downarrow}$
spin-scalar and spin-vector ones, respectively. It is, therefore, natural to call the
corresponding collective variables $X^{+}_{\lambda\mu}(t)$ and 
$X^{-}_{\lambda\mu}(t)$ spin-scalar and spin-vector variables.
The required expressions for $h^{\pm}$, $h^{\uparrow\downarrow}$
and $h^{\downarrow\uparrow}$ are
\begin{eqnarray}\nonumber
&&h_{\tau}^{+}=\frac{p^2}{m}+m\,\omega^2r^2
+12\sum_{\mu}(-1)^{\mu}Z_{2\mu}^{\tau+}(t)\{r\otimes r\}_{2-\mu}
+V_{\tau}^+(\br,t),\\ \nonumber
&&h_{\tau}^-=-\hbar\eta l_0+V_{\tau}^-(\br,t),\quad
h_{\tau}^{\uparrow\downarrow}=-\frac{\hbar}{\sqrt2}\eta l_{-1}+V_{\tau}^{\uparrow\downarrow}(\br,t),
\quad h_{\tau}^{\downarrow\uparrow}=\frac{\hbar}{\sqrt2}\eta l_{1}+V_{\tau}^{\downarrow\uparrow}(\br,t),
\end{eqnarray}
where according to (\ref{Vp})
\begin{eqnarray}
\label{Vss}
V_p^+(\br,t)=-3\frac{\hbar^2}{8}\chi n_p^+(\br,t),\quad
V_p^-(\br,t)=3\frac{\hbar^2}{8}\chi n_p^-(\br,t)+\frac{\hbar^2}{4}\bar\chi n_n^-(\br,t),
\nonumber\\
V_p^{\uparrow\downarrow}(\br,t)=3\frac{\hbar^2}{8}\chi n_p^{\uparrow\downarrow}(\br,t)
+\frac{\hbar^2}{4}\bar\chi n_n^{\uparrow\downarrow}(\br,t),\quad
V_p^{\downarrow\uparrow}(\br,t)=3\frac{\hbar^2}{8}\chi n_p^{\downarrow\uparrow}(\br,t)
+\frac{\hbar^2}{4}\bar\chi n_n^{\downarrow\uparrow}(\br,t)
\end{eqnarray}
and the neutron potentials $V_n^{\varsigma}$ are
obtained by the obvious change of indices $p\leftrightarrow n$.

The integration yields:
\begin{eqnarray}
\label{quadr}
     \dot L^{+}_{\lambda\mu}&=&
\frac{1}{m}P_{\lambda\mu}^{+}-
m\,\omega^2R^{+}_{\lambda \mu}
+12\sqrt5\sum_{j=0}^2\sqrt{2j+1}
\left\{_{2\lambda 1}^{11j}\right\}
\{Z_2^{+}\otimes R_j^{+}\}_{\lambda \mu}
\nonumber\\&&
-i\hbar\frac{\eta}{2}\left[\mu L_{\lambda\mu}^- 
+\sqrt{(\lambda-\mu)(\lambda+\mu+1)}L^{\uparrow\downarrow}_{\lambda\mu+1}+
\sqrt{(\lambda+\mu)(\lambda-\mu+1)}L^{\downarrow\uparrow}_{\lambda\mu-1}\right]
\nonumber\\&&
-\int\!d^3r\left[
\frac{1}{2}n^+\{r\otimes \nabla\}_{\lambda\mu}V^++
\frac{1}{2}n^-\{r\otimes \nabla\}_{\lambda\mu}V^-+
 n^{\downarrow\uparrow}\{r\otimes \nabla\}_{\lambda\mu}V^{\uparrow\downarrow}+
n^{\uparrow\downarrow}\{r\otimes \nabla\}_{\lambda\mu}V^{\downarrow\uparrow}
\right],
\nonumber\\
     \dot L^{-}_{\lambda\mu}&=&
\frac{1}{m}P_{\lambda\mu}^{-}-
m\,\omega^2R^{-}_{\lambda \mu}
+12\sqrt5\sum_{j=0}^2\sqrt{2j+1}\left\{_{2\lambda 1}^{11j}\right\}
\{Z_2^{+}\otimes R_j^{-}\}_{\lambda \mu}-i\hbar\frac{\eta}{2}\mu L_{\lambda\mu}^+
\nonumber\\&&
-\frac{\hbar^2}{2}\eta\delta_{\lambda,1}
\left[\delta_{\mu,-1}F^{\uparrow\downarrow}+\delta_{\mu,1}F^{\downarrow\uparrow}\right]
-\frac{1}{2}\int\!d^3r\left[
n^-\{r\otimes \nabla\}_{\lambda\mu}V^++
n^+\{r\otimes \nabla\}_{\lambda\mu}V^-\right]
\nonumber\\&&
-2\frac{i}{\hbar}\int\! d(\bp,\br)\{r\otimes p\}_{\lambda\mu}
\left[h^{\uparrow\downarrow}f^{\downarrow\uparrow}-h^{\downarrow\uparrow}f^{\uparrow\downarrow}\right],
\nonumber\\
     \dot L^{\uparrow\downarrow}_{\lambda\mu+1}&=&
\frac{1}{m}P_{\lambda\mu+1}^{\uparrow\downarrow}-
m\,\omega^2R^{\uparrow\downarrow}_{\lambda \mu+1}
+12\sqrt5\sum_{j=0}^2\sqrt{2j+1}\left\{_{2\lambda 1}^{11j}\right\}
\{Z_2^{+}\otimes R_j^{\uparrow\downarrow}\}_{\lambda \mu+1}
\nonumber\\&&
-i\hbar\frac{\eta}{4} \sqrt{(\lambda-\mu)(\lambda+\mu+1)}L_{\lambda\mu}^+
+\frac{\hbar^2}{2}\eta\delta_{\lambda,1}
\left[\delta_{\mu,0}F^- +\frac{1}{\sqrt2}\delta_{\mu,-1}F^{\uparrow\downarrow}\right]
\nonumber\\&&
-\frac{1}{2}\int\!d^3r\left[
n^{\uparrow\downarrow}\{r\otimes \nabla\}_{\lambda\mu+1}V^++
n^+\{r\otimes \nabla\}_{\lambda\mu+1}V^{\uparrow\downarrow}\right]
-\frac{i}{\hbar}\int\! d(\bp,\br)\{r\otimes p\}_{\lambda\mu+1}
\left[h^{-}f^{\uparrow\downarrow}-h^{\uparrow\downarrow}f^-\right],
\nonumber
\\
     \dot L^{\downarrow\uparrow}_{\lambda\mu-1}&=&
\frac{1}{m}P_{\lambda\mu-1}^{\downarrow\uparrow}-
m\,\omega^2R^{\downarrow\uparrow}_{\lambda \mu-1}
+12\sqrt5\sum_{j=0}^2\sqrt{2j+1}\left\{_{2\lambda 1}^{11j}\right\}
\{Z_2^{+}\otimes R_j^{\downarrow\uparrow}\}_{\lambda \mu-1}
\nonumber\\&&
-i\hbar\frac{\eta}{4} \sqrt{(\lambda+\mu)(\lambda-\mu+1)}L_{\lambda\mu}^+
+\frac{\hbar^2}{4}\eta\delta_{\lambda,1}
\left[\delta_{\mu,0}F^- -\sqrt2\delta_{\mu,1}F^{\downarrow\uparrow}\right]
\nonumber\\&&
-\frac{1}{2}\int\!d^3r\left[
n^{\downarrow\uparrow}\{r\otimes \nabla\}_{\lambda\mu-1}V^++
n^+\{r\otimes \nabla\}_{\lambda\mu-1}V^{\downarrow\uparrow}\right]
-\frac{i}{\hbar}\int\! d(\bp,\br)\{r\otimes p\}_{\lambda\mu-1}
\left[h^{\downarrow\uparrow}f^- -h^{-}f^{\downarrow\uparrow}\right],
\nonumber
\end{eqnarray}
\begin{eqnarray}
     \dot F^{-}&=&
2\eta \left[L_{1-1}^{\downarrow\uparrow}+L_{11}^{\uparrow\downarrow}\right],
\nonumber\\
     \dot F^{\uparrow\downarrow}&=&
-\eta [L_{1-1}^- -\sqrt2L_{10}^{\uparrow\downarrow}],
\nonumber\\
     \dot F^{\downarrow\uparrow}&=&
-\eta \left[L_{11}^- +\sqrt2L_{10}^{\downarrow\uparrow}\right],
\nonumber\\
     \dot R^{+}_{\lambda\mu}&=&
\frac{2}{m}L^+_{\lambda\mu}
-i\hbar\frac{\eta}{2}\left[\mu R_{\lambda\mu}^- 
+\sqrt{(\lambda-\mu)(\lambda+\mu+1)}R^{\uparrow\downarrow}_{\lambda\mu+1}+
\sqrt{(\lambda+\mu)(\lambda-\mu+1)}R^{\downarrow\uparrow}_{\lambda\mu-1}\right],
\nonumber\\
     \dot R^{-}_{\lambda\mu}&=&
\frac{2}{m}L^-_{\lambda\mu}
-i\hbar\frac{\eta}{2}\mu R_{\lambda\mu}^+
-2\frac{i}{\hbar}\int\! d(\bp,\br)\{r\otimes r\}_{\lambda\mu}
\left[h^{\uparrow\downarrow}f^{\downarrow\uparrow}-h^{\downarrow\uparrow}f^{\uparrow\downarrow}\right],
\nonumber\\
     \dot R^{\uparrow\downarrow}_{\lambda\mu+1}&=&
\frac{2}{m}L^{\uparrow\downarrow}_{\lambda\mu+1}
-i\hbar\frac{\eta}{4} \sqrt{(\lambda-\mu)(\lambda+\mu+1)}R_{\lambda\mu}^+
-\frac{i}{\hbar}\int\! d(\bp,\br)\{r\otimes r\}_{\lambda\mu+1}
\left[h^{-}f^{\uparrow\downarrow}-h^{\uparrow\downarrow}f^-\right],
\nonumber\\
     \dot R^{\downarrow\uparrow}_{\lambda\mu-1}&=&
\frac{2}{m}L^{\downarrow\uparrow}_{\lambda\mu-1}
-i\hbar\frac{\eta}{4} \sqrt{(\lambda+\mu)(\lambda-\mu+1)}R_{\lambda\mu}^+
-\frac{i}{\hbar}\int\! d(\bp,\br)\{r\otimes r\}_{\lambda\mu-1}
\left[h^{\downarrow\uparrow}f^- -h^{-}f^{\downarrow\uparrow}\right],
\nonumber\\
     \dot P^{+}_{\lambda\mu}&=&
-2m\,\omega^2L^+_{\lambda \mu}
+24\sqrt5\sum_{j=0}^2\sqrt{2j+1}\left\{_{2\lambda 1}^{11j}\right\}
\{Z_2^+\otimes L^+_j\}_{\lambda \mu}
\nonumber\\
&&-i\hbar\frac{\eta}{2}\left[\mu P_{\lambda\mu}^- 
+\sqrt{(\lambda-\mu)(\lambda+\mu+1)}P^{\uparrow\downarrow}_{\lambda\mu+1}+
\sqrt{(\lambda+\mu)(\lambda-\mu+1)}P^{\downarrow\uparrow}_{\lambda\mu-1}\right]
\nonumber\\&&
-\int\!d^3r\left[
\{J^+\otimes \nabla\}_{\lambda\mu}V^++
\{J^-\otimes \nabla\}_{\lambda\mu}V^-+
2\{J^{\downarrow\uparrow}\otimes \nabla\}_{\lambda\mu}V^{\uparrow\downarrow}+
2\{J^{\uparrow\downarrow}\otimes \nabla\}_{\lambda\mu}V^{\downarrow\uparrow}
\right],
\nonumber\\
     \dot P^{-}_{\lambda\mu}&=&
-2m\,\omega^2L^-_{\lambda \mu}
+24\sqrt5\sum_{j=0}^2\sqrt{2j+1}\left\{_{2\lambda 1}^{11j}\right\}
\{Z_2^{+}\otimes L^-_j\}_{\lambda \mu}
-i\hbar\frac{\eta}{2}\mu P_{\lambda\mu}^+
\nonumber\\&&
-\int\!d^3r\left[
\{J^-\otimes \nabla\}_{\lambda\mu}V^++
\{J^+\otimes \nabla\}_{\lambda\mu}V^-\right]
-2\frac{i}{\hbar}\int\! d(\bp,\br)\{p\otimes p\}_{\lambda\mu}
\left[h^{\uparrow\downarrow}f^{\downarrow\uparrow}-h^{\downarrow\uparrow}f^{\uparrow\downarrow}\right],
\nonumber\\
     \dot P^{\uparrow\downarrow}_{\lambda\mu+1}&=&
-2m\,\omega^2L^{\uparrow\downarrow}_{\lambda \mu+1}
+24\sqrt5\sum_{j=0}^2\sqrt{2j+1}\left\{_{2\lambda 1}^{11j}\right\}
\{Z_2^{+}\otimes L^{\uparrow\downarrow}_j\}_{\lambda \mu+1}
-i\hbar\frac{\eta}{4} \sqrt{(\lambda-\mu)(\lambda+\mu+1)}P_{\lambda\mu}^+
\nonumber\\&&
-\int\!d^3r\left[
\{J^{\uparrow\downarrow}\otimes \nabla\}_{\lambda\mu+1}V^++
\{J^+\otimes \nabla\}_{\lambda\mu+1}V^{\uparrow\downarrow}\right]
-\frac{i}{\hbar}\int\! d(\bp,\br)\{p\otimes p\}_{\lambda\mu+1}
[h^{-}f^{\uparrow\downarrow}-h^{\uparrow\downarrow}f^-],
\nonumber\\
     \dot P^{\downarrow\uparrow}_{\lambda\mu-1}&=&
-2m\,\omega^2L^{\downarrow\uparrow}_{\lambda \mu-1}
+24\sqrt5\sum_{j=0}^2\sqrt{2j+1}\left\{_{2\lambda 1}^{11j}\right\}
\{Z_2^{+}\otimes L^{\downarrow\uparrow}_j\}_{\lambda \mu-1}
-i\hbar\frac{\eta}{4} \sqrt{(\lambda+\mu)(\lambda-\mu+1)}P_{\lambda\mu}^+
\nonumber\\&&
-\int\!d^3r\left[
\{J^{\downarrow\uparrow}\otimes \nabla\}_{\lambda\mu-1}V^++
\{J^+\otimes \nabla\}_{\lambda\mu-1}V^{\downarrow\uparrow}\right]
\nonumber\\&&
-\frac{i}{\hbar}\int\! d(\bp,\br)\{p\otimes p\}_{\lambda\mu-1}
\left[h^{\downarrow\uparrow}f^- -h^{-}f^{\downarrow\uparrow}\right],
\end{eqnarray}
 where $\left\{_{2\lambda 1}^{11j}\right\}$ is the Wigner
$6j$-symbol and
${\di J_{\nu}^{\varsigma}(\br,t)=\int\frac{d^3p}{(2\pi\hbar)^3}
p_{\nu}f^{\varsigma}(\br,\bp,t)}$ is the current.
For the sake of simplicity the time dependence of tensors is not
written out. It is easy to see that equations (\ref{quadr}) are
nonlinear due to quadrupole-quadrupole and spin-spin interactions.
We will solve them in the small amplitude approximation, by linearizing
the equations. This procedure helps also to solve another problem: to
represent the integral terms in (\ref{quadr}) as the linear combination
of collective variables (\ref{Varis}), that allows to close the whole 
set of  equations (\ref{quadr}). The detailed analysis of the integral
terms is given in the appendix \ref{AppA}.

We are interested in the scissors mode with quantum number
$K^{\pi}=1^+$. Therefore, we only need the part of dynamic equations 
with $\mu=1$.

\subsection{Linearized equations ($\mu=1$), isovector, isoscalar}

 Writing all variables as a sum of their
equilibrium value plus a small deviation
$$R_{\lambda\mu}(t)=R_{\lambda\mu}^{\rm eq}+\R_{\lambda\mu}(t),\quad
P_{\lambda\mu}(t)=P_{\lambda\mu}^{\rm eq}+\P_{\lambda\mu}(t),\quad
L_{\lambda\mu}(t)=L_{\lambda\mu}^{\rm eq}+\L_{\lambda\mu}(t)$$
and neglecting quadratic deviations, one obtains the linearized
equations. Naturally one needs to know the equilibrium values of all
variables.
Evident equilibrium conditions for an axially symmetric nucleus are:
\begin{equation}
R^{+}_{2\pm1}(eq)=R^{+}_{2\pm2}(eq)=0,\quad R^{+}_{20}(eq)\neq0.
\label{equi1}
\end{equation}
It is obvious that all ground state properties of the system of spin
up nucleons are identical to the ones of the system of nucleons with spin
down. Therefore
\begin{equation}
R^{-}_{\lambda\mu}(eq)=P^{-}_{\lambda\mu}(eq)=L^{-}_{\lambda\mu}(eq)=0.
\label{equi2}
\end{equation}
We also will suppose 
\begin{equation}
L^{+}_{\lambda\mu}(eq)=L^{\uparrow\downarrow}_{\lambda\mu}(eq)=L^{\downarrow\uparrow}_{\lambda\mu}(eq)=0
\quad\mbox{ and }\quad
R^{\uparrow\downarrow}_{\lambda\mu}(eq)=R^{\downarrow\uparrow}_{\lambda\mu}(eq)=0.
\label{equi3}
\end{equation}
Let us recall
that all variables and equilibrium quantities $R^{+}_{\lambda 0}(eq)$
and $Z^{+}_{20}(eq)$ in (\ref{quadr}) have isospin indices 
$\tau=n,\,p$. All the difference 
between neutron and proton systems is contained in the mean field 
quantities $Z^{\tau+}_{20}(eq)$ and $V_{\tau}^{\varsigma}$, which
are different for neutrons and protons (see eq.~(\ref{Z2mu}) and~(\ref{Vss})).

  It is convenient to rewrite the dynamical equations in terms
of isovector and isoscalar variables
$$R_{\lambda\mu}=R_{\lambda\mu}^{n}+R_{\lambda\mu}^{p},\quad
P_{\lambda\mu}=P_{\lambda\mu}^{n}+P_{\lambda\mu}^{p},\quad
L_{\lambda\mu}=L_{\lambda\mu}^{n}+L_{\lambda\mu}^{p},$$
$$\bar R_{\lambda\mu}=R_{\lambda\mu}^{n}-R_{\lambda\mu}^{p},\quad
\bar P_{\lambda\mu}=P_{\lambda\mu}^{n}-P_{\lambda\mu}^{p},\quad
\bar L_{\lambda\mu}=L_{\lambda\mu}^{n}-L_{\lambda\mu}^{p}.$$
It also is natural to define isovector and isoscalar strength constants
$\kappa_1=\frac{1}{2}(\kappa-\bar\kappa)$ and
$\kappa_0=\frac{1}{2}(\kappa+\bar\kappa)$ connected by the relation
$\kappa_1=\alpha\kappa_0$ \cite{BaSc}.
Then the equations for the neutron and proton systems are transformed
into isovector and isoscalar ones. Supposing that all equilibrium
characteristics of the proton system are equal to that of the neutron
system one decouples isovector and isoscalar equations. This 
approximations looks rather crude, nevertheless the possible 
corrections to it are very small, being of the order 
$(\frac{N-Z}{A})^2$.
With the help of the above equilibrium relations one arrives at the 
following final set of equations for the isovector system:
\begin{eqnarray}
\label{iv}
     \dot {\bar\L}^{+}_{21}&=&
\frac{1}{m}\bar\P_{21}^{+}-
\left[m\,\omega^2
-4\sqrt3\alpha\kappa_0R_{00}^{\rm eq}
+\sqrt6(1+\alpha)\kappa_0 R_{20}^{\rm eq}\right]\bar\R^{+}_{21}
-i\hbar\frac{\eta}{2}\left[\bar\L_{21}^-
+2\bar\L^{\uparrow\downarrow}_{22}+
\sqrt6\bar\L^{\downarrow\uparrow}_{20}\right],
\nonumber\\
     \dot {\bar\L}^{-}_{21}&=&
\frac{1}{m}\bar\P_{21}^{-}
-\left[m\,\omega^2+\sqrt6\kappa_0 R_{20}^{\rm eq}
-\frac{\sqrt{3}}{20}\hbar^2 
\left( \chi-\frac{\bar\chi}{3} \right)
\left(\frac{I_1}{a_0^2}+\frac{I_1}{a_1^2}\right)\left(\frac{a_1^2}{{\cal A}_2}-\frac{a_0^2}{{\cal A}_1}\right)
\right]\bar\R^{-}_{21}
-i\hbar\frac{\eta}{2}\bar\L_{21}^+,
\nonumber\\
     \dot {\bar\L}^{\uparrow\downarrow}_{22}&=&
\frac{1}{m}\bar\P_{22}^{\uparrow\downarrow}-
\left[m\,\omega^2-2\sqrt6\kappa_0R_{20}^{\rm eq}
-\frac{\sqrt{3}}{5}\hbar^2 
\left( \chi-\frac{\bar\chi}{3} \right)\frac{I_1}{{\cal A}_2}
\right]\bar\R^{\uparrow\downarrow}_{22}
-i\hbar\frac{\eta}{2}\bar\L_{21}^+,
\nonumber\\
     \dot {\bar\L}^{\downarrow\uparrow}_{20}&=&
\frac{1}{m}\bar\P_{20}^{\downarrow\uparrow}-
\left[m\,\omega^2
+2\sqrt6\kappa_0 R_{20}^{\rm eq}\right]\bar\R^{\downarrow\uparrow}_{20}
+\frac{2}{\sqrt3}\kappa_0 R_{20}^{\rm eq}\,\bar\R^{\downarrow\uparrow}_{00}
-i\hbar\frac{\eta}{2}\sqrt{\frac{3}{2}}\bar\L_{21}^+ 
\nonumber\\
&&+\frac{\sqrt{3}}{15}\hbar^2 
\left( \chi-\frac{\bar\chi}{3} \right)I_1 \,\frac
{\left({\cal A}_1-2{\cal A}_2\right)\bar \R_{20}^{\downarrow\uparrow}+
\sqrt2\left({\cal A}_1+{\cal A}_2\right) \bar \R_{00}^{\downarrow\uparrow}
}{{\cal A}_1{\cal A}_2},
\nonumber\\
     \dot {\bar\L}^{+}_{11}&=&
-3\sqrt6(1-\alpha)\kappa_0 R_{20}^{\rm eq}\,\bar\R^{+}_{21}
-i\hbar\frac{\eta}{2}\left[\bar\L_{11}^- 
+\sqrt2\bar\L^{\downarrow\uparrow}_{10}\right],
\nonumber\\
     \dot {\bar\L}^{-}_{11}&=&
-\left[3\sqrt6\kappa_0 R_{20}^{\rm eq}
-\frac{\sqrt{3}}{20}\hbar^2 
\left( \chi-\frac{\bar\chi}{3} \right)
\left(\frac{I_1}{a_0^2}-\frac{I_1}{a_1^2}\right)\left(\frac{a_1^2}{{\cal A}_2}-\frac{a_0^2}{{\cal A}_1}\right)
\right]\bar\R^{-}_{21}
-\hbar\frac{\eta}{2}\left[i\bar\L_{11}^+
+\hbar\bar F^{\downarrow\uparrow}\right],
\nonumber\\
     \dot {\bar\L}^{\downarrow\uparrow}_{10}&=&
-\hbar\frac{\eta}{2\sqrt2}\left[i\bar\L_{11}^+
+\hbar\bar F^{\downarrow\uparrow}\right],
\nonumber\\
     \dot {\bar F}^{\downarrow\uparrow}&=&
-\eta\left[\bar\L_{11}^- +\sqrt2\bar\L^{\downarrow\uparrow}_{10}\right],
\nonumber
\end{eqnarray}\begin{eqnarray}
     \dot {\bar\R}^{+}_{21}&=&
\frac{2}{m}\bar\L_{21}^{+}
-i\hbar\frac{\eta}{2}\left[\bar\R_{21}^-
+2\bar\R^{\uparrow\downarrow}_{22}+
\sqrt6\bar\R^{\downarrow\uparrow}_{20}\right],
\nonumber\\
     \dot {\bar\R}^{-}_{21}&=&
\frac{2}{m}\bar\L_{21}^{-}
-i\hbar\frac{\eta}{2}\bar\R_{21}^+,
\nonumber\\
     \dot {\bar\R}^{\uparrow\downarrow}_{22}&=&
\frac{2}{m}\bar\L_{22}^{\uparrow\downarrow}
-i\hbar\frac{\eta}{2}\bar\R_{21}^+,
\nonumber\\
     \dot {\bar\R}^{\downarrow\uparrow}_{20}&=&
\frac{2}{m}\bar\L_{20}^{\downarrow\uparrow}
-i\hbar\frac{\eta}{2}\sqrt{\frac{3}{2}}\bar\R_{21}^+,
\nonumber\\
     \dot {\bar\P}^{+}_{21}&=&
-2\left[m\,\omega^2+\sqrt6\kappa_0 R_{20}^{\rm eq}\right]\bar\L^{+}_{21}
+6\sqrt6\kappa_0 R_{20}^{\rm eq}\bar\L^{+}_{11}
-i\hbar\frac{\eta}{2}\left[\bar\P_{21}^- 
+2\bar\P^{\uparrow\downarrow}_{22}+\sqrt6\bar\P^{\downarrow\uparrow}_{20}\right]
\nonumber\\
&&+\frac{3\sqrt{3}}{4}\hbar^2 
\chi \frac{I_2}{{\cal A}_1{\cal A}_2}
\left[\left({\cal A}_1-{\cal A}_2\right) \bar\L_{21}^{+} +
\left({\cal A}_1+{\cal A}_2\right) \bar\L_{11}^{+}\right],
\nonumber\\
     \dot {\bar\P}^{-}_{21}&=&
-2\left[m\,\omega^2+\sqrt6\kappa_0 R_{20}^{\rm eq}\right]\bar\L^{-}_{21}
+6\sqrt6\kappa_0 R_{20}^{\rm eq}\bar\L^{-}_{11}
-i\hbar\frac{\eta}{2}\bar\P_{21}^{+}
\nonumber\\
&&+\frac{3\sqrt{3}}{4}\hbar^2 
\chi \frac{I_2}{{\cal A}_1{\cal A}_2}
\left[\left({\cal A}_1-{\cal A}_2\right)\bar\L_{21}^{-} +
\left({\cal A}_1+{\cal A}_2\right) \bar\L_{11}^{-}\right],
\nonumber\\
     \dot {\bar\P}^{\uparrow\downarrow}_{22}&=&
-\left[2m\,\omega^2-4\sqrt6\kappa_0 R_{20}^{\rm eq}
-\frac{3\sqrt{3}}{2}\hbar^2 
\chi \frac{I_2}{{\cal A}_2}
\right]\bar\L^{\uparrow\downarrow}_{22}
-i\hbar\frac{\eta}{2}\bar\P_{21}^{+},
\nonumber\\
     \dot {\bar\P}^{\downarrow\uparrow}_{20}&=&
-\left[2m\,\omega^2+4\sqrt6\kappa_0 R_{20}^{\rm eq}\right]\bar\L^{\downarrow\uparrow}_{20}
+8\sqrt3\kappa_0 R_{20}^{\rm eq}\bar\L^{\downarrow\uparrow}_{00}
-i\hbar\frac{\eta}{2}\sqrt{\frac{3}{2}}\bar\P_{21}^{+}
\nonumber\\
&&+\frac{\sqrt{3}}{2}\hbar^2 
\chi \frac{I_2}{{\cal A}_1{\cal A}_2}
\left[\left({\cal A}_1-2{\cal A}_2\right)\bar\L_{20}^{\downarrow\uparrow}+
\sqrt2\left({\cal A}_1+{\cal A}_2\right) \bar\L_{00}^{\downarrow\uparrow}
\right],
\nonumber\\
     \dot {\bar\L}^{\downarrow\uparrow}_{00}&=&
\frac{1}{m}\bar\P_{00}^{\downarrow\uparrow}-m\,\omega^2\bar\R^{\downarrow\uparrow}_{00}
+4\sqrt3\kappa_0 R_{20}^{\rm eq}\,\bar\R^{\downarrow\uparrow}_{20}
\nonumber\\
&&+\frac{1}{2\sqrt{3}}\hbar^2 
\left[\left( \chi-\frac{\bar\chi}{3} \right)I_1-\frac{9}{4}\chi I_2\right]
\frac{
\left(2{\cal A}_1-{\cal A}_2\right)\bar \R_{00}^{\downarrow\uparrow}+
\sqrt2\left({\cal A}_1+{\cal A}_2\right) \bar \R_{20}^{\downarrow\uparrow}
}{{\cal A}_1{\cal A}_2},
\nonumber\\
     \dot {\bar\R}^{\downarrow\uparrow}_{00}&=&
\frac{2}{m}\bar\L_{00}^{\downarrow\uparrow},
\\
\nonumber
     \dot {\bar\P}^{\downarrow\uparrow}_{00}&=&
-2m\,\omega^2\bar\L^{\downarrow\uparrow}_{00}
+8\sqrt3\kappa_0 R_{20}^{\rm eq}\,\bar\L^{\downarrow\uparrow}_{20}
+\frac{\sqrt{3}}{2}\hbar^2 
\chi I_2
\left[\left(\frac2{\A_2}-\frac1{\A_1}\right)\bar\L_{00}^{\downarrow\uparrow}+
\sqrt2\left(\frac1{\A_2}+\frac1{\A_1}\right)\bar\L_{20}^{\downarrow\uparrow}
\right],
\end{eqnarray}
where ${\cal A}_1,\ {\cal A}_2$ are defined in appendix \ref{AppB}, $\kappa_0=-m\bar\omega^2/(4Q_{00})$~\cite{BrMt}
with $\bar\omega^2=\omega^2/\left(1+\frac23\delta\right)$,
$\displaystyle a_{-1} = a_1 = R_0\left( \frac{1-(2/3)\delta}{1+(4/3)\delta} \right)^{1/6}$ and
$\displaystyle a_0 = R_0\left( \frac{1-(2/3)\delta}{1+(4/3)\delta} \right)^{-1/3}$ 
are semiaxes of ellipsoid by which the shape of nucleus is approximated, $\delta$ -- deformation parameter,
$R_0=1.2A^{1/3}$~fm.
\begin{eqnarray}\label{Int_ss} \nonumber
I_1=\frac{\pi}{4}\int\limits_{-\infty}^{+\infty}dr\, r^4\left(\frac{\partial n^+(r)}{\partial r}\right)^2,
\
I_2=\frac{\pi}{4}\int\limits_{-\infty}^{+\infty}dr\, r^2 n^+(r)^2,\quad 
n^+(r)=n_p^{+}+n_n^{+}= \frac{n_0}{1+{\rm e}^{\frac{r-R_0}{a}}}.
\end{eqnarray}

The isoscalar set of equations is easily obtained from~(\ref{iv}) by 
taking $\alpha=1$ and replacing $\bar\chi \to -\bar\chi$.

\section{Discussion and interpretation of the results}

The energies and excitation probabilities obtained by the solution of
the {\bf isovector} set of equations (\ref{iv}) are given in Table \ref{tab1}. 
The used spin-spin interaction is repulsive, the values of its
strength constants being taken from the paper \cite{Moya}, where the 
notation $\chi=K_s/A,\,\bar\chi=q\chi$ was introduced.
The results without spin-spin interaction (variant I) are compared
with those performed with 
two sets of constants $K_s,\, q$ (variants II, III). 
The first set of constants (variant II) was extracted by the authors of~\cite{Moya} 
from Skyrme forces following the standard procedure, the residual interaction 
being defined in terms of second derivatives of the Hamiltonian density 
$H(\rho)$ with respect to the one-body densities~$\rho$.
Different variants of Skyrme forces produce different strength constants of 
spin-spin interaction. The most consistent results are obtained with 
SG1, SG2 \cite{Giai} and Sk3 \cite{Floc} forces. We use here the 
spin-spin constants extracted from Sk3 force.
Another set of constants (variant III) was also found by the authors 
of~\cite{Moya} phenomenologically in the calculations with a Woods-Saxon
potential, when there is not any self-consistency between 
the mean field and the residual interaction. 
We tentatively will use it, because in our case also there is no self-consistency.
The strength of the spin-orbit interaction 
is taken from~\cite{Solov}.

\begin{table*}
\caption{\label{tab1}
 Isovector energies and excitation probabilities 
of $^{164}$Er. Deformation parameter $\delta=0.25$, spin-orbit constant $\eta=0.36$ MeV. 
Spin-spin interaction constants are: I -- $K_s=0$ MeV;
II -- $K_s=92$ MeV, $q=-0.8$;
 III~--~$K_s=200$~MeV, $q=-0.5$.
Quantum numbers (including indices $\varsigma = +,\,-,\,
\uparrow\downarrow,\,\downarrow\uparrow$) of variables responsible for 
the generation of the present level are shown in the first column. 
For example: $(1,1)^-$ -- spin scissors, $(1,1)^+$ -- conventional scissors, etc..
}
\begin{ruledtabular}
\begin{tabular}{lrrrrrrrrr}
\multicolumn{1}{c}{ $(\lambda,\mu)^\varsigma$} &   
\multicolumn{3}{c}{ $E_{\rm iv}$, MeV}      &
\multicolumn{3}{c}{ $B(M1),\ \mu_N^2$}      &
\multicolumn{3}{c }{ $B(E2),  \ B_W$}        \\
\cline{2-10} &
\multicolumn{1}{c}{ I } & \multicolumn{1}{c}{ II} & \multicolumn{1}{c}{ III} &
\multicolumn{1}{c}{ I } & \multicolumn{1}{c}{ II} & \multicolumn{1}{c}{ III} &
\multicolumn{1}{c}{ I } & \multicolumn{1}{c}{ II} & \multicolumn{1}{c }{ III} \\ 
\hline
 (1,1)$^-$                    & ~1.61  & ~2.02  & ~2.34 & ~3.54 & ~5.44 & ~7.91 & ~0.12 & ~0.36 & ~0.82 \\   
 (1,1)$^+$                    & ~2.18  & ~2.45  & ~2.76 & ~5.33 & ~4.48 & ~2.98 & ~1.02 & ~1.23 & ~1.26 \\ 
 (0,0)$^{\downarrow\uparrow}$ & 12.80  & 16.81  & 20.02 & ~0.01 & ~0.01 & ~0.04 & ~0.04 & ~0.13 & ~0.52 \\  
 (2,1)$^-$                    & 14.50  & 18.52  & 21.90 & ~0.01 & ~0.02 & ~0.34 & ~0.03 & ~0.13 & ~4.29 \\  
 (2,2)$^{\uparrow\downarrow}$ & 16.18  & 20.61  & 24.56 & ~0.02 & ~0.23 & ~0.03 & ~0.18 & ~3.09 & ~0.44 \\   
 (2,0)$^{\downarrow\uparrow}$ & 16.20  & 22.65  & 27.67 & ~0    & ~0.03 & ~0    & ~0    & ~0.39 & ~0.02 \\  
 (2,1)$^+$                    & 20.59  & 21.49  & 22.42 & ~2.78 & ~2.19 & ~1.77 & 35.45 & 30.47 & 27.43 \\
 (1,0)$^{\downarrow\uparrow}$ & ~0.26i & ~0.26i &~0.26i & -5.4i & -5.4i & -5.4i &~~0i   &~~0i   & ~0i   \\
\end{tabular}
\end{ruledtabular}
\end{table*}

One can see from Table \ref{tab1} that the spin-spin interaction does not change
the qualitative picture of the positions of the excitations described 
in \cite{BaMo}. It pushes all levels
up proportionally to its strength (20-30\% in the case II and 40-60\%
in the case III) without changing their order. The most interesting
result concerns the relative B(M1) values of the two low lying scissors modes,
namely the spin scissors $(1,1)^-$ and the conventional (orbital) 
scissors $(1,1)^+$ mode. As can be noticed, the spin-spin interaction 
strongly redistributes M1 strength in the favour of the spin scissors 
mode. We tentatively want to link this fact to the recent experimental
finding in isotopes of Th and Pa \cite{Siem}. The authors have studied 
deuteron and $^3$He-induced reactions on $^{232}$Th and found in the 
residual nuclei $^{231,232,233}$Th and $^{232,233}$Pa
"an unexpectedly strong integrated strength of $B(M1)=11-15~\mu_N^2$ in the 
$E_\gamma=1.0-3.5$~MeV region". The $B(M1)$ force in most nuclei shows 
evident splitting into two Lorentzians. "Typically, the experimental
splitting is $\Delta\omega_{M1}\sim 0.7$~MeV, and the ratio of the 
strengths between the lower and upper resonance components is
$B_L/B_U\sim 2$". (Note a misprint in that paper: it is written 
erroneously $B_2/B_1\sim 2$ whereas it should be $B_1/B_2\sim 2$. 
To avoid misunderstanding, we write here $B_L$ instead of $B_1$ 
and $B_U$ instead of $B_2$.) The authors have tried to explain the 
splitting by a $\gamma$-deformation. To describe the observed value of
$\Delta\omega_{M1}$ the deformation $\gamma\sim 15^\circ$ is required, 
that leads to the ratio $B_L/B_U\sim 0.7$ in an obvious contradiction 
with experiment. The authors conclude that "the splitting may be 
due to other mechanisms". In this sense, we tentatively may argue as
follows. On one side, theory~\cite{Gogny} and experiment~\cite{Korten}
give zero value of $\gamma$-deformation for $^{233}$Th. On the other 
side, it is easy to see that our theory suggests the required mechanism. 
The calculations performed for $^{233}$Th give 
$\Delta\omega_{M1}\sim 0.32$~MeV and $B_L/B_U\sim 1.6$ for the first 
variant of the spin-spin interaction and 
$\Delta\omega_{M1}\sim 0.28$~MeV and $B_L/B_U\sim 4.1$ for second one 
in reasonable agreement with experimental values. The inclusion of 
pair correlations will affect our results, but one may speculate that 
the agreement between the theory and experiment will be conserved at 
least qualitatively.

The energies and excitation probabilities obtained by the solution of
the {\bf isoscalar} set of equations (\ref{iv}) are displayed in the 
Table \ref{tab2}. The general picture of the influence of the spin-spin 
interaction here is quite close to that observed in the isovector
case. The only difference is the low lying mode marked by $(1,1)^+$
which is practically insensitive to the spin-spin interaction.
The negligibly small
negative $B(M1)$ value of spin scissors appears undoubtedly due to the
lack of the self consistency in our calculations.
In ref \cite{Siem} the assignment of the resonances to be of isovector type 
is only tentative based on the assumption that at such low energies there 
is no  collective mode other than the isovector scissors mode. However, 
from \cite{Siem} one cannot exclude that also an isoscalar spin scissors mode 
is mixed in. From our analysis we see that the isoscalar spin scissors 
where all nucleons with spin up counter-rotate with respect the ones of spin up 
comes more or less at the same energy as the isovector scissors. So it 
would be very important for the future to pin down precisely the quantum 
numbers of the resonances.

\begin{table*}
\caption{\label{tab2} The same as in Table \ref{tab1}, but for isoscalar excitations.}
\begin{ruledtabular}
\begin{tabular}{lrrrrrrrrr}
\multicolumn{1}{c}{ $(\lambda,\mu)^\varsigma$} &   
\multicolumn{3}{c}{ $E_{\rm is}$, MeV}      &
\multicolumn{3}{c}{ $B(M1),\ \mu_N^2$}      &
\multicolumn{3}{c }{ $B(E2),  \ B_W$}        \\
\cline{2-10} &
\multicolumn{1}{c}{ I } & \multicolumn{1}{c}{ II} & \multicolumn{1}{c}{ III} &
\multicolumn{1}{c}{ I } & \multicolumn{1}{c}{ II} & \multicolumn{1}{c}{ III} &
\multicolumn{1}{c}{ I } & \multicolumn{1}{c}{ II} & \multicolumn{1}{c }{ III} \\ 
\hline
 (1,1)$^-$                    & ~1.73  & ~2.04  & ~2.40  & -0.07 & -0.05 & ~0    & ~~1.12 & ~~0.65 & ~~0.39 \\
 (1,1)$^+$                    & ~0.39  & ~0.37  & ~0.37  & ~0.24 & ~0.24 & ~0.24 & 117.2  & 117.9  & 118.3  \\
 (0,0)$^{\downarrow\uparrow}$ & 12.83  & 15.59  & 18.72  & ~0    & ~0    & ~0    & ~~0.66 & ~~0.31 & ~~0.15 \\
 (2,1)$^-$                    & 14.51  & 17.40  & 20.65  & ~0    & ~0    & ~0    & ~~0.12 & ~~0.06 & ~~0.03 \\
 (2,2)$^{\uparrow\downarrow}$ & 16.20  & 19.43  & 23.09  & ~0    & ~0    & ~0    & ~~0    & ~~0.07 & ~~0.04 \\ 
 (2,0)$^{\downarrow\uparrow}$ & 16.22  & 20.09  & 24.80  & ~0    & ~0    & ~0    & ~~0.20 & ~~0.02 & ~~0.01 \\
 (2,1)$^+$                    & 10.28  & 11.92  & 13.60  & ~0    & ~0    & ~0    & ~66.50 & ~57.78 & ~50.87 \\
 (1,0)$^{\downarrow\uparrow}$ & ~0.20i & ~0.20i & ~0.20i &-0.1i  &-0.1i  &-0.1i  & ~30.0i & ~29.8i & ~30.3i \\
\end{tabular}
\end{ruledtabular}
\end{table*}
\end{widetext}

Let us discuss in more detail the nature of the predicted excitations. 
As one sees, the generalization of
the WFM method by including spin dynamics allowed one to
reveal a variety of new types of nuclear collective motion involving 
spin degrees of freedom. Two isovector and two isoscalar low
lying eigenfrequencies and five isovector and five isoscalar high 
lying eigenfrequencies have been found.

Three low lying levels correspond to the excitation of new types of 
modes. For example the isovector level marked by $(1,1)^-$ describes
rotational oscillations of nucleons with the spin projection "up"
with respect of nucleons with the spin projection "down", i.e. one
can talk of a nuclear spin scissors mode. Having in mind that this 
excitation is an isovector one, we can see that the resulting motion 
looks rather complex -- proton spin scissors counter-rotates with 
respect to the neutron spin scissors. Thus the experimentally 
observed group of 1$^+$ peaks in the interval 2-4 MeV, associated 
usually with the nuclear scissors mode, in reality consists of the 
excitations of the "spin" scissors mode together with the conventional
\cite{Heyd} scissors mode (the level $(1,1)^+$ in our case).
The isoscalar level $(1,1)^-$ describes the real spin 
scissors mode: all spin up nucleons (protons together with neutrons)
oscillate rotationally out of phase with all spin down 
nucleons.

Such excitations were, undoubtedly, produced implicitly
by other methods (e.g. RPA~\cite{Heyd,Pena,Moya,Oster}), but they never 
were analysed in such terms. It is interesting to note, for example,
that in~\cite{Pena} the scissors mode was analyzed in 
so-called spin and orbital components. Roughly speaking there are two groups of 
states corresponding to these two types of components, not completely 
dissimilar to our finding. Whereas the nature of the orbital, i.e. conventional 
scissors is quite clear, the authors did not analyze the character of their 
states which consist of the spin component. It can be speculated that those 
spin components just correspond to the isovector spin scissors mode discussed 
in our work here. It would be interesting to study whether our suggestion is 
correct or not. This could for example be done in analyzing the current 
patterns.

One more new low lying mode (isoscalar, marked by $(1,1)^+$) is 
generated by the relative motion of the orbital angular momentum and 
spin of the nucleus (they can change their absolute values and 
directions keeping the total spin unchanged).

In order to complete the picture of the low-lying states, it is important to 
discuss the state which is slightly imaginary. Let us first state that the 
nature of this state has nothing to do with neither spin scissors nor with 
conventional scissors. It can namely be seen from the structure of our 
equations that this state corresponds to a spin flip induced by the spin-orbit 
potential. Such a state is of purely quantal character and it cannot be hoped 
that we can accurately describe it with our WFM approach restricting 
the consideration by second order moments only.
 For its correct treatment, we certainly should consider higher moments
like fourth order moments, for instance. The spin-orbit 
potential is the only term in our theory which couples  the second order 
moments to the fourth order ones. As mentioned, we decoupled the system in 
neglecting the fourth order moments. Therefore, it is no surprise that this 
particular spin flip mode is not well described. Nevertheless, one may try to 
better understand the origin of this mode almost at zero energy. For this, we 
make the following approximation of our diagonalisation procedure to get the 
eight eigenvalues listed in Table \ref{tab1}. We neglect in (\ref{iv}) all 
couplings between the set of variables 
$X^+_{\lambda\mu}, X^-_{\lambda\mu}$ and the set of variables
$X^{\uparrow\downarrow}_{\lambda\mu}, X^{\downarrow\uparrow}_{\lambda\mu}$.
To this end in the dynamical equations for 
$X^+_{\lambda\mu}, X^-_{\lambda\mu}$ we omit all terms containing
$X^{\uparrow\downarrow}_{\lambda\mu}, X^{\downarrow\uparrow}_{\lambda\mu}$
and in the dynamical equations for
$X^{\uparrow\downarrow}_{\lambda\mu}, X^{\downarrow\uparrow}_{\lambda\mu}$
we omit all terms containing $X^+_{\lambda\mu}, X^-_{\lambda\mu}$.
In such a way we get two independent sets of dynamical equations. The
first one (for $X^+_{\lambda\mu}, X^-_{\lambda\mu}$) was already 
studied in \cite{BaMo}, where we have found that such approximation
gives satisfactory (in comparison with the exact solution) results but
must be used cautiously because of the problems with the angular 
momentum conservation. The second set of equations (for 
$X^{\uparrow\downarrow}_{\lambda\mu}, X^{\downarrow\uparrow}_{\lambda\mu}$)
splits into three independent subsets. Two of them were already analyzed
in \cite{BaMo} (it turns out that these subsets can be obtained also
in the limit $\eta\to 0$, which was studied there), where it
was shown that the results of approximate calculations are very close
to that of exact calculations, i.e. the coupling between the respective
variables 
$X^{\uparrow\downarrow}_{\lambda\mu}, X^{\downarrow\uparrow}_{\lambda\mu}$
and $X^+_{\lambda\mu}, X^-_{\lambda\mu}$ is very weak. The only new
subset of equations reads:
\begin{eqnarray}
\label{LF}
     \dot {\bar\L}^{\downarrow\uparrow}_{10}&=&
-\hbar^2\frac{\eta}{2\sqrt2}
\bar F^{\downarrow\uparrow},
\nonumber\\
     \dot {\bar F}^{\downarrow\uparrow}&=&
-\eta\sqrt2\bar\L^{\downarrow\uparrow}_{10}.
\end{eqnarray}
The solution of these equations is  ${\di E =i\frac{\hbar}{\sqrt2}\eta = i\,0.255}$
what practically coincides with 
the number of the full diagonalisation. So the non-zero (purely imaginary) 
value of this root only comes from the fact that z-component of orbital angular 
momentum is not conserved (only total spin J is conserved). However, the 
violation of the conservation of orbital angular momentum is very small as can 
be seen from the numbers. In any case, we see that this spin flip state has 
nothing to do with neither the spin scissors nor with the conventional 
scissors.

Two high lying excitations of a new nature are found. They are
marked by $(2,1)^-$ and following the paper \cite{Oster} can be 
called spin-vector giant quadrupole
resonances. The isovector one corresponds to the following quadrupole
motion: the proton system oscillates out of phase with the neutron 
system, whereas inside of each system spin up nucleons oscillate out 
of phase with spin down nucleons. The respective isoscalar resonance 
describes out of phase oscillations of all spin up nucleons
(protons together with neutrons) with respect of all spin down 
nucleons.

Six high lying modes can be interpreted 
as spin-flip giant monopole (marked by $(0,0)^{\downarrow\uparrow}$)
and quadrupole (marked by $(2,0)^{\downarrow\uparrow}$ and 
$(2,2)^{\uparrow\downarrow}$) resonances.

It is a pertinent place to make following citation from the review by
F. Osterfeld~\cite{Oster}: {\it "Similar oscillations to
those in isospin space are also possible in spin space. Nucleons with
spin up and spin down may move either in phase (spin-scalar S=0 modes)
or out of phase (spin-vector S=1 modes). The latter class of states is
also referred to as spin excitations or spin-flip transitions."} 
On account of our results in this work, the latter statement
that all spin excitations 
are of spin-flip nature should be modified. We predict in this paper the 
existence of spin excitations 
of non spin-flip nature -- the isovector and isoscalar spin scissors
and the isovector and isoscalar spin-vector GQR!

\section{Concluding remarks}

The inclusion of spin-spin interaction does not change qualitatively
the picture concerning the spectrum of the spin modes found in \cite{BaMo}. It 
pushes all levels up without changing their order. However, it strongly 
redistributes M1 strength 
between the conventional and spin scissors mode in the favour of the last one.
Our calculations did not fully confirm the expectations mentioned
in the introduction, namely that essentially only the low lying part of the 
spectrum will be strongly influenced by the spin-spin force. Nevertheless 
our results turned out to be very useful, because
they demonstrate that the common
effect of spin-spin interaction and pair correlations are able
to push a substantial part of the M1 force out of the area of the 
conventional scissors mode what is required for the reasonable
agreement with experimental data.

In this respect, we should mention that we did not include pairing in 
this work. 
Inclusion of pairing would have complicated the formalism quite a 
bit. It shall be worked out in the future.
We here wanted to study the features of spin dynamics in a most 
transparent way staying, however, somewhat on the qualitative side. 
That is why we did not try to discuss in detail possible relations with 
experiment or to compare with the results of other theories. 
Nevertheless we mentioned the quite recent experimental
work~\cite{Siem}, where for the two low lying magnetic states a stronger 
B(M1) transition for the lower state with respect to the higher one was 
found. A tentative explanation in terms of a slight triaxial deformation 
in~\cite{Siem} failed. However, our theory can naturally predict such a scenario with
a non vanishing spin-spin force.
It would indeed be very exciting, if the results of~\cite{Siem} had 
already discovered the isovector spin scissors mode. However, much 
deeper experimental and theoretical results must be obtained before a 
firm conclusion on this point is possible.

In the light of the above results, the study of spin excitations with 
pairing included, will be the natural continuation of this work. 
Pairing is important for a quantitative description of the 
conventional scissors mode. The same is expected for the novel spin 
scissors mode discussed here. The effect of pairing generally is to push 
up the spectrum in energy. Therefore, as just mentioned, it can be expected 
that the results come into better agreement with experiment.

\begin{acknowledgments}
The fruitful discussions with D. Pena Arteaga, Nguen Van Giai, 
V. O. Nesterenko, A. I. Vdovin and J. N. Wilson are gratefully acknowledged.
\end{acknowledgments}

\appendix
\section{ }
\label{AppA}

All derivations of this section will be done in the approximation of
spherical symmetry. The inclusion of deformation makes the calculations
more cumbersome without changing the final conclusions.
Let us consider, as an example, the integral 
$$I_h=\int\! d(\bp,\br)\{r\otimes p\}_{\lambda\mu}
[h^{\uparrow\downarrow}f^{\downarrow\uparrow}-h^{\downarrow\uparrow}f^{\uparrow\downarrow}].$$
It can be divided in two parts corresponding to the contributions
of spin-orbital and spin-spin potentials: $I_h=I_{so}+I_{ss},$
where
$$I_{so}=-\frac{\hbar}{\sqrt2}\eta\int\! d(\bp,\br)\{r\otimes p\}_{\lambda\mu}
[l_{-1}f^{\downarrow\uparrow} + l_{1}f^{\uparrow\downarrow}],$$
$$I_{ss}=\int\! d(\bp,\br)\{r\otimes p\}_{\lambda\mu}
[V_{\tau}^{\uparrow\downarrow}f^{\downarrow\uparrow}-V_{\tau}^{\downarrow\uparrow}f^{\uparrow\downarrow}],$$
$V_{\tau}^{ss'}$ being defined in (\ref{Vss}).
It is easy to see that the integral $I_{so}$ generate moments of 
fourth order. 
According to the rules of the WFM method \cite{Bal} this integral 
is neglected.

Let us analyze the integral $I_{ss}$ (to be definite, for protons).
In this case
$$V_p^{\uparrow\downarrow}(\br)=3\frac{\hbar^2}{8}\chi n_p^{\uparrow\downarrow}(\br)
+\frac{\hbar^2}{4}\bar\chi n_n^{\uparrow\downarrow}(\br),$$
$$V_p^{\downarrow\uparrow}(\br)=3\frac{\hbar^2}{8}\chi n_p^{\downarrow\uparrow}(\br)
+\frac{\hbar^2}{4}\bar\chi n_n^{\downarrow\uparrow}(\br).$$
It is seen that $I_{ss}$ is split into four terms of 
identical structure, so it will be sufficient to analyze in detail 
only one part. For example
\begin{widetext}
\begin{eqnarray}
\label{Iss4}
I_{ss4}
=\int\! d(\bp,\br)\{r\otimes p\}_{\lambda\mu}n^{\downarrow\uparrow}f^{\uparrow\downarrow}
=\int\! d^3r\{r\otimes J^{\uparrow\downarrow}\}_{\lambda\mu}n^{\downarrow\uparrow}
=\sum_{\nu,\alpha}C^{\lambda\mu}_{1\nu,1\alpha}\int\! d^3r 
r_{\nu}J^{\uparrow\downarrow}_{\alpha}n^{\downarrow\uparrow},
\end{eqnarray}
where ${ J_{\alpha}^{\uparrow\downarrow}(\br,t)=\int\frac{d^3p}{(2\pi\hbar)^3}
p_{\alpha}f^{\uparrow\downarrow}(\br,\bp,t)}$.
The variation of this integral reads
\begin{eqnarray}
\label{Var4}
\delta I_{ss4}
=\sum_{\nu,\alpha}C^{\lambda\mu}_{1\nu,1\alpha}\int\! d\,^3r r_{\nu}
\left[n^{\downarrow\uparrow}(eq)\delta J^{\uparrow\downarrow}_{\alpha}+
J^{\uparrow\downarrow}_{\alpha}(eq)\delta n^{\downarrow\uparrow}\right].
\end{eqnarray}
It is necessary to represent this integral in terms of the collective
variables (\ref{Varis}). This problem can not be solved exactly, so 
we will use the approximation suggested in \cite{Bal} and expand the 
density and current variations as a series (see appendix \ref{AppB}).

Let us consider the second part of integral (\ref{Var4}). With the 
help of formula (\ref{Varn1}) we find
\begin{eqnarray}
\label{Var42}
&&I_{2}
\equiv\sum_{\nu,\alpha}C^{\lambda\mu}_{1\nu,1\alpha}\int\! d^3r\, r_{\nu}
J^{\uparrow\downarrow}_{\alpha}(eq)\delta n^{\downarrow\uparrow}
\nonumber\\
&&=-\sum_{\nu,\alpha}C^{\lambda\mu}_{1\nu,1\alpha}\int\! d^3r\, r_{\nu}
J^{\uparrow\downarrow}_{\alpha}(eq)
\sum_{\beta}(-1)^{\beta}\left\{
N^{\downarrow\uparrow}_{\beta,-\beta}(t)n^+
+\sum_{\gamma}(-1)^{\gamma}N^{\downarrow\uparrow}_{\beta,\gamma}(t)\frac{1}{r}
\frac{\partial n^+}{\partial r}r_{-\beta}r_{-\gamma}
\right\}.\qquad
\end{eqnarray}
Let us analyze at first the more simple part of this expression:
\begin{eqnarray}
\label{Var422}
I_{2,1}\equiv -\sum_{\beta}(-1)^{\beta}N^{\downarrow\uparrow}_{\beta,-\beta}(t)
\int\! d^3r\, 
\sum_{\nu,\alpha}C^{\lambda\mu}_{1\nu,1\alpha}
r_{\nu}J^{\uparrow\downarrow}_{\alpha}(eq)n^+=
-\sum_{\beta}(-1)^{\beta}N^{\downarrow\uparrow}_{\beta,-\beta}X_{\lambda\mu}.
\end{eqnarray}
We are interested in the value of $\mu=1$, therefore it is necessary 
to analyze two possibilities: $\lambda=1$ and $\lambda=2$.

In the case $\lambda=1,\,\mu=1$ we have
\begin{eqnarray}
\label{Lam11}
X_{11}\equiv \int\! d^3r\, n^+\sum_{\nu,\alpha}C^{11}_{1\nu,1\alpha}
r_{\nu}J^{\uparrow\downarrow}_{\alpha}(eq)
=\int\! d^3r\, n^+\frac{1}{\sqrt2}\left[r_1J^{\uparrow\downarrow}_0(eq)-r_0J^{\uparrow\downarrow}_1(eq)\right].
\end{eqnarray}
By definition
\begin{eqnarray}
\label{Jeq}
J^{ss'}_{\nu}=
\int\frac{d^3p}{(2\pi\hbar)^3}p_{\nu}f^{ss'}(\br,\bp)
=-\frac{i\hbar}{2}\Big[(\nabla_{\nu}-\nabla'_{\nu})
\rho(\br,s;\br's')\Big]_{\br'=\br}
\nonumber\\
=\frac{i\hbar}{2}\sum_k v_k^2[
\phi_k(\br,s)\nabla_{\nu}\phi_k^*(\br,s')
-\phi_k^*(\br,s')\nabla_{\nu}\phi_k(\br,s)],
\end{eqnarray}
where $k\equiv n,l,j,m$ is a set of oscillator quantum numbers, $v_k^2$ are
occupation numbers, and
\begin{eqnarray}
\label{phi}
\phi_{nljm}(\br,s)
=\R_{nl}(r)\sum_{\Lambda,\sigma}
C^{j m}_{l\Lambda,\frac{1}{2}\sigma}
 Y_{l\Lambda}(\theta,\phi)\chi_{\frac{1}{2}\sigma}(s)
=\R_{nlj}(r)C^{jm}_{lm-s,\frac12 s}Y_{lm-s}(\theta,\phi)
\end{eqnarray}
are single particle wave functions, $\chi_{\frac{1}{2}\sigma}(s)=\delta_{\sigma,s}$
being spin functions.
Inserting (\ref{Jeq}) into (\ref{Lam11}) one finds
\begin{eqnarray}
\label{L11}
X_{11}=\frac{i\hbar}{2}\frac{1}{\sqrt2}\sum_{nljm} v_{nljm}^2
\int\! d^3r\, n^+(r)\R_{nlj}^2(r)C^{jm}_{l\Lambda,\frac12 \frac12}
C^{jm}_{l\Lambda',\frac12 -\frac12}\left[Y_{l\Lambda}(r_1\nabla_0-r_0\nabla_1)Y_{l\Lambda'}^*\right.
\nonumber\\
\left.-Y_{l\Lambda'}^*(r_1\nabla_0-r_0\nabla_1)Y_{l\Lambda}\right]
\end{eqnarray}
with $\Lambda=m-\frac12$ and $\Lambda'=m+\frac12$. Remembering the definition
(\ref{lxyz}) of the angular momentum $\hat l_1=\hbar(r_0\nabla_1-r_1\nabla_0)$
and using the relation \cite{Var} 
$\hat l_{\pm1}Y_{l\Lambda}=\mp\frac1{\sqrt2}\sqrt{(l\mp\Lambda)(l\pm\Lambda+1)}Y_{l\Lambda\pm1}$
one transforms (\ref{L11}) into
\begin{eqnarray}
\label{L11a}
X_{11}
=-\frac{i\hbar}{2}\frac{1}{\sqrt2}\sum_{nljm} v_{nljm}^2
\int\!dr n^+(r)r^2\R_{nlj}^2(r)C^{jm}_{l\Lambda,\frac12 \frac12}
C^{jm}_{l\Lambda',\frac12 -\frac12}\frac2{\sqrt2}\sqrt{(l-\Lambda)(l+\Lambda+1)}
\nonumber\\
=-i\hbar\sum_{nl}\sum_{m=\frac12}^{|l-\frac12|}
\frac{[(l+\frac12)^2-m^2]}{2l+1}
\int\!dr n^+(r)r^2\left[v_{nll+\frac12m}^2\R_{nll+\frac12}^2(r)
-v_{nl|l-\frac12|m}^2\R_{nl|l-\frac12|}^2(r)\right].
\end{eqnarray}
As it is seen, the value of this integral is determined by the difference
of the wave functions of spin-orbital partners $(v\R)_{nll+\frac12m}^2
-(v\R)_{nl|l-\frac12|m}^2$, which is usually very small, so we will
neglect it. The only remarkable contribution can appear in the vicinity
of the Fermi surface, where some spin-orbital partners with 
$j=l+\frac12$ and $j=|l-\frac12|$ can be disposed on different sides of
the Fermi surface. In reality such situation happens very frequently,
nevertheless we will not take into account this effect, because the 
values of the corresponding integrals are considerably smaller than
$R_{20}(eq)$, the typical input parameter of our model.

Let us consider now the integral $I_{2,1}$ (formula (\ref{Var422}))
for the case $\lambda=2,\,\mu=1$. We have
\begin{eqnarray}
\label{Lam21}
X_{21}\equiv \int\! d^3r n^+\sum_{\nu,\alpha}C^{21}_{1\nu,1\alpha}
r_{\nu}J^{\uparrow\downarrow}_{\alpha}(eq)
=\int\! d^3r n^+C^{21}_{11,10}\left[r_1J^{\uparrow\downarrow}_0(eq)+r_0J^{\uparrow\downarrow}_1(eq)\right].
\end{eqnarray}
With the help of formulae (\ref{Jeq}) and (\ref{phi}) one can show by
simple algebraic transformations that 
\begin{eqnarray}
\label{Lam21'}
\int\! d\Omega \,r_1J^{\uparrow\downarrow}_0(eq)=-\int\! d\Omega \,r_0J^{\uparrow\downarrow}_1(eq),
\end{eqnarray}
where $\int\! d\Omega$ means the integration over angles. As a result
$X_{21}=0$.

Let us consider the second, more complicated, part of 
integral $I_2$:
\begin{eqnarray}
\label{Var421}
I_{2,2}
&=&-\sum_{\beta,\gamma}(-1)^{\beta+\gamma}N^{\downarrow\uparrow}_{-\beta,-\gamma}(t)
\sum_{\nu,\alpha}C^{\lambda\mu}_{1\nu,1\alpha}\int\! d\,^3r r_{\nu}
J^{\uparrow\downarrow}_{\alpha}(eq)
\frac{1}{r}\frac{\partial n^+}{\partial r}r_{\beta}r_{\gamma}
\nonumber\\
&=&-\sum_{\beta,\gamma}(-1)^{\beta+\gamma}
N^{\downarrow\uparrow}_{-\beta,-\gamma}(t)X_{\lambda\mu}'(\beta,\gamma).
\end{eqnarray}
The case $\lambda=1,\,\mu=1$:
\begin{eqnarray}
\label{Lam11'}
&&X_{11}'(\beta,\gamma)
=\frac{1}{\sqrt2}\int\! d^3r\, \frac{1}{r}\frac{\partial n^+}{\partial r}
\left[r_1J^{\uparrow\downarrow}_0(eq)-r_0J^{\uparrow\downarrow}_1(eq)\right]r_{\beta}r_{\gamma}
\nonumber\\
&&=-\frac{i\hbar}{4}\sum_{nljm} v_{nljm}^2
\int\! d^3r\, \frac{1}{r}\frac{\partial n^+}{\partial r}
\R_{nlj}^2(r)C^{jm}_{l\Lambda,\frac12 \frac12}
C^{jm}_{l\Lambda',\frac12 -\frac12}
\sqrt{(l-\Lambda)(l+\Lambda+1)}
\left[Y_{l\Lambda}Y_{l\Lambda}^*
+Y_{l\Lambda'}^*Y_{l\Lambda'}\right]
r_{\beta}r_{\gamma}.\qquad
\end{eqnarray}
The angular part of this integral is
\begin{eqnarray}
\label{Ome}
\int\!\! d\Omega
\left[Y_{l\Lambda}Y_{l\Lambda}^*
+Y_{l\Lambda'}^*Y_{l\Lambda'}\right]r_{\beta}r_{\gamma}
=\sum_{L,M}C^{LM}_{1\beta,1\gamma}
\int\!\! d\Omega
\left[Y_{l\Lambda}Y_{l\Lambda}^*
+Y_{l\Lambda'}^*Y_{l\Lambda'}\right]\{r\otimes r\}_{LM}
\nonumber\\
=-\frac{2}{\sqrt3}\,r^2 C^{00}_{1\beta,1\gamma}+\sqrt{\frac{8\pi}{15}}r^2
\sum_{M}C^{2M}_{1\beta,1\gamma}
\int\!\! d\Omega 
\left[Y_{l\Lambda}Y_{l\Lambda}^*+Y_{l\Lambda'}^*Y_{l\Lambda'}\right]Y_{2M}
\nonumber\\
=\frac{2}{3}\,r^2\delta_{\gamma,-\beta}\left\{1-\sqrt{\frac{5}{2}}\, C^{l0}_{l0,20}
C^{1\beta}_{1\beta,20}\left[C^{l\Lambda}_{l\Lambda,2M}+C^{l\Lambda'}_{l\Lambda',2M}\right]\right\}.
\end{eqnarray}
Therefore
\begin{eqnarray}
\label{I21}
X_{11}'(\beta,\gamma)
=
-\frac{i\hbar}{6}
\delta_{\gamma,-\beta}
\int\! dr\, \frac{\partial n^+(r)}{\partial r}\,r^3
\sum_{nljm}
\left\{1-\sqrt{\frac{5}{2}}\,C^{1\beta}_{1\beta,20}
C^{l0}_{l0,20}\left[C^{l\Lambda}_{l\Lambda,20}+C^{l\Lambda'}_{l\Lambda',20}\right]
\right\}\times
\nonumber\\
 v_{nljm}^2\R_{nlj}^2(r)
C^{jm}_{l\Lambda,\frac12 \frac12}
C^{jm}_{l\Lambda',\frac12 -\frac12}
\sqrt{(l-\Lambda)(l+\Lambda+1)}
\nonumber\\
=
-\frac{i\hbar}{3}
\delta_{\gamma,-\beta}
\sum_{nl}
\left\{1-\sqrt{\frac{5}{2}}\,C^{1\beta}_{1\beta,20}
C^{l0}_{l0,20}\left[C^{l\Lambda}_{l\Lambda,20}+C^{l\Lambda'}_{l\Lambda',20}\right]
\right\}\times
\nonumber\\
\sum_{m=\frac12}^{|l-\frac12|}
\frac{[(l+\frac12)^2-m^2]}{2l+1}
\int\! dr \frac{\partial n^+(r)}{\partial r}r^3
\left[v_{nll+\frac12m}^2\R_{nll+\frac12}^2(r)
-v_{nl|l-\frac12|m}^2\R_{nl|l-\frac12|}^2(r)\right].
\end{eqnarray}
One sees that, exactly as in formula (\ref{L11a}), the value of this
integral is determined by the difference of the wave functions of 
spin-orbital partners $(v\R)_{nll+\frac12m}^2-(v\R)_{nl|l-\frac12|m}^2$ near the Fermi surface, 
so it can be omitted together with $X_{11}$ following the same arguments.

The case $\lambda=2, \mu=1$ can be analyzed in full analogy with
formulae (\ref{Lam21},\ref{Lam21'}) that allows us to take $X_{21}'=0$.

So, we have shown that the integral $I_2$ can be approximated by zero.
Let us consider now the first part of the integral (\ref{Var4}):
\begin{eqnarray}
\label{Var41}
I_1
&=&\sum_{\nu,\alpha}C^{\lambda\mu}_{1\nu,1\alpha}\int\! d^3r r_{\nu}
n^{\downarrow\uparrow}(eq)\delta J^{\uparrow\downarrow}_{\alpha}
=\sum_{\nu,\alpha}C^{\lambda\mu}_{1\nu,1\alpha}\int\! d^3r r_{\nu}
n^{\downarrow\uparrow}(eq)n^+(r)
\sum_{\gamma}(-1)^{\gamma}K^{\uparrow\downarrow}_{\alpha,-\gamma}(t)r_{\gamma}
\nonumber\\
&=&\sum_{\nu,\alpha}C^{\lambda\mu}_{1\nu,1\alpha}
\sum_{\gamma}(-1)^{\gamma}K^{\uparrow\downarrow}_{\alpha,-\gamma}(t)
\int\! d^3r 
n^{\downarrow\uparrow}(eq)n^+(r)
\sum_{L,M}C^{LM}_{1\nu,1\gamma}
\{r\otimes r\}_{LM}.
\end{eqnarray}
This integral can be estimated in the approximation of constant density
$n^+(r)=n_0$. Then
\begin{eqnarray}
\label{Var41a}
I_1
=n_0\sum_{\nu,\alpha}C^{\lambda\mu}_{1\nu,1\alpha}
\sum_{\gamma}(-1)^{\gamma}K^{\uparrow\downarrow}_{\alpha,-\gamma}(t)
\sum_{L,M}C^{LM}_{1\nu,1\gamma}
R^{\downarrow\uparrow}_{LM}(eq)=0.
\end{eqnarray}
It is easy to show, that $R^{\downarrow\uparrow}_{LM}(eq)=0.$ Let us consider, for 
example, the case with $L=2$:
\begin{eqnarray}
\label{Rd}
R^{\downarrow\uparrow}_{2M}
=\int\! d(\bp,\br) \{r\otimes r\}_{2M}f^{\downarrow\uparrow}(\br,\bp)
=\int\! d^3r \{r\otimes r\}_{2M}n^{\downarrow\uparrow}(\br)
=\sqrt{\frac{8\pi}{15}}\int\! d^3r r^2Y_{2M}n^{\downarrow\uparrow}(\br).
\end{eqnarray}
By definition
\begin{eqnarray}
\label{neq}
n^{ss'}(\br)=
\int\frac{d^3p}{(2\pi\hbar)^3}f^{ss'}(\br,\bp)
=\sum_k v_k^2
\phi_k(\br,s)\phi_k^*(\br,s')
\end{eqnarray}
with $\phi_k$ defined in (\ref{phi}). Therefore
\begin{eqnarray}
\label{Rd'}
R^{\downarrow\uparrow}_{2M}
=\sqrt{\frac{8\pi}{15}}
\int\!d^3r r^2Y_{2M}\sum_{nljm} v_{nljm}^2\R_{nlj}^2(r)
C^{jm}_{l\Lambda',\frac12 -\frac12}
C^{jm}_{l\Lambda,\frac12 \frac12}
Y_{l\Lambda'}Y_{l\Lambda}^*
\nonumber\\
=\sqrt{\frac{2}{3}}\sum_{nljm} v_{nljm}^2
\int\!dr r^4\R_{nlj}^2(r)
C^{jm}_{l\Lambda,\frac12 \frac12}
C^{jm}_{l\Lambda',\frac12 -\frac12}
C^{l0}_{20,l0}
C^{l\Lambda}_{2M,l\Lambda'}=0,
\end{eqnarray}
where $\Lambda=m-\frac12$ and $\Lambda'=m+\frac12$.
The zero is obtained due to summation over $m$. Really, the
product $C^{jm}_{l\Lambda,\frac12 \frac12}
C^{jm}_{l\Lambda',\frac12 -\frac12}=\pm\frac{\sqrt{(l+\frac12)^2-m^2}}{2l+1}$
(for $j=l\pm\frac12$) does not depend on the sign of $m$, whereas the
Clebsh-Gordan coefficient $C^{l\Lambda}_{2M,l\Lambda'}$ changes its
sign together with $m$.

Summarizing, we have demonstrated that $I_1+I_2\simeq 0$, hence one can 
neglect the contribution of the integrals $I_h$ in the equations of motion.

$\bullet$ It is necessary to analyze also the integrals with the weight
$\{p\otimes p\}_{\lambda\mu}$:
$$I'_h=\int\! d(\bp,\br)\{p\otimes p\}_{\lambda\mu}
\left[h^{\uparrow\downarrow}f^{\downarrow\uparrow}-h^{\downarrow\uparrow}f^{\uparrow\downarrow}\right]=I'_{so}
+I'_{ss}.$$
Again we neglect the contribution of the spin-orbital part $I'_{so}$,
which generates fourth order moments. For the spin-spin contribution, we have
\begin{eqnarray}
\label{I'ss4}
I'_{ss4}
=\int\! d(\bp,\br)\{p\otimes p\}_{\lambda\mu}n^{\downarrow\uparrow}(\br,t)f^{\uparrow\downarrow}(\br,\bp,t)
=\int\! d^3r\Pi^{\uparrow\downarrow}_{\lambda\mu}(\br,t)n^{\downarrow\uparrow}(\br,t),
\end{eqnarray}
where $\Pi^{\uparrow\downarrow}_{\lambda\mu}(\br,t)=\int\frac{d^3p}{(2\pi\hbar)^3}
\{p\otimes p\}_{\lambda\mu}f^{\uparrow\downarrow}(\br,\bp,t)$ is the pressure tensor.
The variation of this integral reads:
\begin{eqnarray}
\label{VarI'}
\delta I'_{ss4}
=\int\! d^3r\left[n^{\downarrow\uparrow}(eq)\delta\Pi^{\uparrow\downarrow}_{\lambda\mu}(\br,t)
+\Pi^{\uparrow\downarrow}_{\lambda\mu}(eq)\delta n^{\downarrow\uparrow}(\br,t)\right].
\end{eqnarray}
The pressure tensor variation is defined in appendix \ref{AppB}.
With formula (\ref{VarP1}) one finds for the first part of 
(\ref{VarI'}):
\begin{eqnarray}
\label{VarI'1}
I'_1
=\int\! d^3r n^{\downarrow\uparrow}(eq)\delta\Pi^{\uparrow\downarrow}_{\lambda\mu}(\br,t)
\simeq
T^{\uparrow\downarrow}_{\lambda\mu}(t)\int\! d^3r n^{\downarrow\uparrow}(eq)n^+(r)
\simeq
T^{\uparrow\downarrow}_{\lambda\mu}(t)n_0\int\! d^3r n^{\downarrow\uparrow}(eq)=0.
\end{eqnarray}
The last equality follows obviously from the definition of $n^{\downarrow\uparrow}$ (\ref{neq}).

The second part of (\ref{VarI'}) reads:
\begin{eqnarray}
\label{VarI'2}
I'_2
&=&\int\! d^3r\Pi^{\uparrow\downarrow}_{\lambda\mu}(eq)\delta n^{\downarrow\uparrow}(r,t)
\nonumber\\
&=&-\sum_{\beta}(-1)^{\beta}\int\! d^3r\Pi^{\uparrow\downarrow}_{\lambda\mu}(eq)
\left\{
N^{\downarrow\uparrow}_{\beta,-\beta}(t)n^+
+\sum_{\gamma}(-1)^{\gamma}N^{\downarrow\uparrow}_{\beta,\gamma}(t)\frac{1}{r}
\frac{\partial n^+}{\partial r}r_{-\beta}r_{-\gamma}
\right\}.
\end{eqnarray}
Let us consider at first the simpler part of this integral
\begin{eqnarray}
\label{I'22}
-\sum_{\beta}(-1)^{\beta}N^{\downarrow\uparrow}_{\beta,-\beta}(t)
\int\! d^3r\Pi^{\uparrow\downarrow}_{\lambda\mu}(eq)n^+(r).
\end{eqnarray}
The value of the last integral is determined by the angular structure
of the function $\Pi^{\uparrow\downarrow}_{\lambda\mu}(\br)$. We are interested in
$\lambda=2, \mu=1$. By definition
\begin{eqnarray}
\label{Pieq}
\Pi^{\uparrow\downarrow}_{21}(\br)=\int\frac{d^3p}{(2\pi\hbar)^3}
\{p\otimes p\}_{21}f^{\uparrow\downarrow}(\br,\bp)
=\sum_{\nu,\sigma}C^{21}_{1\nu,1\sigma}
\int\frac{d^3p}{(2\pi\hbar)^3}
p_{\nu}p_{\sigma}f^{\uparrow\downarrow}(\br,\bp)
\nonumber\\
=2C^{21}_{11,10}
\int\frac{d^3p}{(2\pi\hbar)^3}
p_{1}p_{0}f^{\uparrow\downarrow}(\br,\bp)
=-\frac{\hbar^2}{2\sqrt2}\left[(\nabla'_1-\nabla_1)(\nabla'_0-\nabla_0)
\rho(\br'\uparrow,\br\downarrow)\right]_{\br'=\br}
\nonumber\\
=-\frac{\hbar^2}{2\sqrt2}\sum_k v_k^2
\left\{
[\nabla_1\nabla_0\phi_k(\br,\uparrow)]\phi_k^*(\br,\downarrow)
-[\nabla_1\phi_k(\br,\uparrow)][\nabla_0\phi_k^*(\br,\downarrow)]
\right.\nonumber\\ \left.
-[\nabla_0\phi_k(\br,\uparrow)][\nabla_1\phi_k^*(\br,\downarrow)]
+\phi_k(\br,\uparrow)[\nabla_1\nabla_0\phi_k^*(\br,\downarrow)]
\right\}
\end{eqnarray}
with $\phi_k$ being defined by (\ref{phi}).
Taking into account formulae \cite{Var}
$$\nabla_{\pm1}Y_{l\lambda}=
-\sqrt{\frac{(l\pm\Lambda+1)(l\pm\Lambda+2)}{2(2l+1)(2l+3)}}
\frac lrY_{l+1,\Lambda\pm1}
-\sqrt{\frac{(l\mp\Lambda-1)(l\mp\Lambda)}{2(2l-1)(2l+1)}}
\frac{l+1}rY_{l-1,\Lambda\pm1},
$$
$$\nabla_{0}Y_{l\lambda}=
-\sqrt{\frac{(l+1)^2-\Lambda^2}{(2l+1)(2l+3)}}
\frac lrY_{l+1,\Lambda}
+\sqrt{\frac{l^2-\Lambda^2}{(2l-1)(2l+1)}}
\frac{l+1}rY_{l-1,\Lambda}
$$
one finds that 
\begin{eqnarray}
\label{I'Pi}
\int\! d^3r\Pi^{\uparrow\downarrow}_{\lambda\mu}(eq)n^+(r)=
\hbar^2\sum_{nljm} v_{nljm}^2
\int\!dr n^+(r)\R_{nlj}^2(r)
(\delta_{j,l+\frac12}-\delta_{j,l-\frac12})
\frac{l(l+1)[(l+\frac12)^2-m^2]}{(2l+3)(2l+1)(2l-1)}m=0\qquad
\end{eqnarray}
due to summation over $m$. The more complicated part of the integral 
(\ref{VarI'2}) is calculated in a similar way with the same result,
hence $I'_2=0$.

So, we have shown that $I'_1+I'_2\simeq 0$, therefore one can neglect 
by the contribution of integrals $I'_h$ (together with $I_h$) into 
equations of motion.

$\bullet$ And finally, just a few words about the integrals with the 
weight $\{r\otimes r\}_{\lambda\mu}$:
$$I''_h=\int\! d(\bp,\br)\{r\otimes r\}_{\lambda\mu}
\left[h^{\uparrow\downarrow}f^{\downarrow\uparrow}-h^{\downarrow\uparrow}f^{\uparrow\downarrow}\right]=I''_{so}
+I''_{ss}.$$
The spin-orbital part $I''_{so}$ is neglected and for the spin-spin part we have
\begin{eqnarray}
\label{I"ss4}
I''_{ss4}
=\int\! d(\bp,\br)\{r\otimes r\}_{\lambda\mu}n^{\downarrow\uparrow}(\br,t)
f^{\uparrow\downarrow}(\br,\bp,t)
=\int\! d^3r\{r\otimes r\}_{\lambda\mu}n^{\downarrow\uparrow}(\br,t)n^{\uparrow\downarrow}(\br,t).
\end{eqnarray}
The variation of this integral reads:
\begin{eqnarray}
\label{VarI"}
\delta I''_{ss4}
=\int\! d^3r\{r\otimes r\}_{\lambda\mu}[n^{\downarrow\uparrow}(eq)\delta n^{\uparrow\downarrow}(\br,t)
+n^{\uparrow\downarrow}(eq)\delta n^{\downarrow\uparrow}(\br,t)].
\end{eqnarray}
With the help of formulae (\ref{neq}) and (\ref{Varn1}) the subsequent
analysis becomes quite similar to that of the integral (\ref{Var41})
with the same result, i.e. $I''_h\simeq 0.$

$\bullet$ The integrals 
$\int\! d(\bp,\br)W_{\lambda\mu}
\left[h^-f^{\downarrow\uparrow}-h^{\downarrow\uparrow}f^-\right]$ and
$\int\! d(\bp,\br)W_{\lambda\mu}
\left[h^-f^{\uparrow\downarrow}-h^{\uparrow\downarrow}f^-\right]$,
where $W_{\lambda\mu}$ is any of the above mentioned weights, can
be analyzed in an analogous way with the same result.

\section{ }
\label{AppB}

According to the approximation suggested in \cite{Bal}, the variations
of density, current, and pressure tensor are expanded as the series
\begin{eqnarray}
\label{Varn}
\delta n^{\varsigma}(\br,t)
&=&-\sum_{\beta}(-1)^{\beta}\nabla_{-\beta}\left\{n^+(\br)\left[N^{\varsigma}_{\beta}(t)
+\sum_{\gamma}(-1)^{\gamma}N^{\varsigma}_{\beta,\gamma}(t)r_{-\gamma}
\right.\right.\nonumber\\ 
&&\left.\left.+\sum_{\lambda',\mu'}(-1)^{\mu'}N^{\varsigma}_{\beta,\lambda'\mu'}(t)
\{r\otimes r\}_{\lambda'-\mu'}+...\right]\right\},
\end{eqnarray}
\begin{eqnarray}
\label{VarJ}
\delta J^{\varsigma}_{\beta}(\br,t)
=n^+(\br)\left[K^{\varsigma}_{\beta}(t)
+\sum_{\gamma}(-1)^{\gamma}K^{\varsigma}_{\beta,-\gamma}(t)r_{\gamma}
+\sum_{\lambda',\mu'}(-1)^{\mu'}K^{\varsigma}_{\beta,\lambda'-\mu'}(t)
\{r\otimes r\}_{\lambda'\mu'}+...\right],
\end{eqnarray}
\begin{eqnarray}
\label{VarP}
\delta \Pi^{\varsigma}_{\lambda\mu}(\br,t)
=n^+(\br)\left[T^{\varsigma}_{\lambda\mu}(t)
+\sum_{\gamma}(-1)^{\gamma}T^{\varsigma}_{\lambda\mu,-\gamma}(t)r_{\gamma}
+\sum_{\lambda',\mu'}(-1)^{\mu'}T^{\varsigma}_{\lambda\mu,\lambda'-\mu'}(t)
\{r\otimes r\}_{\lambda'\mu'}+...\right].
\end{eqnarray}
Putting these series into the integrals (\ref{Var4}, \ref{VarI'}), one discovers
immediately that all terms containing expansion coefficients $N,\,K,\,T$
with odd numbers of indices disappear due to axial symmetry. Furthermore,
we truncate these series omitting all terms generating higher (than 
second) order moments. So, finally the following expressions are used:
\begin{eqnarray}
\label{Varn1}
\delta n^{\varsigma}(\br,t)
&\simeq &-\sum_{\beta}(-1)^{\beta}\nabla_{-\beta}\left\{n^+(\br)
\sum_{\gamma}(-1)^{\gamma}N^{\varsigma}_{\beta,\gamma}(t)r_{-\gamma}\right\}
\nonumber\\
&=&-\sum_{\beta}(-1)^{\beta}\left\{
N^{\varsigma}_{\beta,-\beta}(t)n^+
+\sum_{\gamma}(-1)^{\gamma}N^{\varsigma}_{\beta,\gamma}(t)\frac{1}{r}
\frac{\partial n^+}{\partial r}r_{-\beta}r_{-\gamma}
\right\},
\end{eqnarray}
\begin{eqnarray}
\label{VarJ1}
\delta J^{\varsigma}_{\beta}(\br,t)
\simeq n^+(\br)
\sum_{\gamma}(-1)^{\gamma}K^{\varsigma}_{\beta,-\gamma}(t)r_{\gamma}
\end{eqnarray}
and
\begin{eqnarray}
\label{VarP1}
\delta \Pi^{\varsigma}_{\lambda\mu}(\br,t)
\simeq n^+(\br)T^{\varsigma}_{\lambda\mu}(t).
\end{eqnarray}
The coefficients $N^{\varsigma}_{\beta,\gamma}(t)$ and $K^{\varsigma}_{\beta,-\gamma}(t)$
are connected by the linear relations with collective variables $\R^{\varsigma}_{\lambda\mu}(t)$
and $\L^{\varsigma}_{\lambda\mu}(t)$ respectively.
\begin{eqnarray}
\R_{\lambda\mu}^\varsigma
=\int\! d^3r \{r\otimes r\}_{\lambda\mu}\delta n^\varsigma(\br)
=\frac{2}{\sqrt3}\left[{\cal A}_1 C_{1\mu,10}^{\lambda\mu}N_{\mu,0}^\varsigma-
{\cal A}_2 \left(C_{1\mu+1,1-1}^{\lambda\mu}N_{\mu+1,-1}^\varsigma+
C_{1\mu-1,11}^{\lambda\mu}N_{\mu-1,1}^\varsigma\right)\right],
\end{eqnarray}
where
\begin{eqnarray}
\label{CA} 
{\cal A}_1=\sqrt2\, R_{20}^{\rm eq}-R_{00}^{\rm eq}=\frac{Q_{00}}{\sqrt3}\left(1+\frac{4}{3}\delta\right),\quad
{\cal A}_2= R_{20}^{\rm eq}/\sqrt2+R_{00}^{\rm eq}=-\frac{Q_{00}}{\sqrt3}\left(1-\frac{2}{3}\delta\right),
\end{eqnarray}
$R_{20}=Q_{20}/\sqrt6$, $R_{00}=-Q_{00}/\sqrt3$, $Q_{20}=\frac43\delta Q_{00}$, $Q_{00}=A<r^2>=\frac35 AR_0^2$.
\begin{eqnarray}
\label{G} 
\nonumber&&N_{-1,-1}^\varsigma=-{\frac{\sqrt3\,\R_{2-2}^\varsigma}{2{\cal A}_2}},\quad
N_{-1,0}^\varsigma={\frac{\sqrt6\,\R_{2-1}^\varsigma}{4{\cal A}_1}},\quad
N_{-1,1}^\varsigma=-{\frac{\R_{00}^\varsigma+\R_{20}^\varsigma/\sqrt2}{2{\cal A}_2}},\\
\nonumber&&N_{0,-1}^\varsigma=-{\frac{\sqrt6\,\R_{2-1}^\varsigma}{4{\cal A}_2}},\quad
N_{0,0}^\varsigma={\frac {\sqrt {2}\R_{2,0}^\varsigma-\R_{0,0}^\varsigma}{2{\cal A}_1}},\quad
N_{0,1}^\varsigma=-{\frac{\sqrt6\,\R_{21}^\varsigma}{4{\cal A}_2}},\\
&&N_{1,-1}^\varsigma=N_{-1,1}^\varsigma,\quad
N_{1,0}^\varsigma={\frac{\sqrt6\,\R_{21}^\varsigma}{4{\cal A}_1}},\quad
N_{1,1}^\varsigma=-{\frac{\sqrt3\,\R_{22}^\varsigma}{2{\cal A}_2}}.
\end{eqnarray}
\begin{eqnarray}
\L_{\lambda,\mu}^\varsigma
=\int\! d^3r \{r\otimes \delta J^\varsigma\}_{\lambda\mu}
=\frac{1}{\sqrt3}(-1)^\lambda 
\left[{\cal A}_1 C_{1\mu,10}^{\lambda\mu}K_{\mu,0}^\varsigma-
{\cal A}_2 \left(C_{1\mu+1,1-1}^{\lambda\mu}K_{\mu+1,-1}^\varsigma+
C_{1\mu-1,11}^{\lambda\mu}K_{\mu-1,1}^\varsigma\right)\right].\qquad
\end{eqnarray}
\begin{eqnarray}
\label{K} 
\nonumber&&K_{-1,-1}^\varsigma=-{\frac{\sqrt3\,\L_{2-2}^\varsigma}{{\cal A}_2}},
\quad
K_{-1,0}^\varsigma={\frac{\sqrt3\,(\L_{1-1}^\varsigma+\L_{2-1}^\varsigma)}{\sqrt2\,{\cal A}_1}},
\quad
K_{-1,1}^\varsigma=-{\frac{\sqrt3\, \L_{10}^\varsigma+\L_{20}^\varsigma+\sqrt2\, \L_{00}^\varsigma}
{\sqrt2\, {\cal A}_2}},\\
\nonumber&&K_{0,-1}^\varsigma={\frac{\sqrt3\,(\L_{1-1}^\varsigma-\L_{2-1}^\varsigma)}{\sqrt2\,{\cal A}_2}},
\quad
K_{0,0}^\varsigma={\frac {\sqrt {2}\L_{2,0}^\varsigma-\L_{0,0}^\varsigma}{{\cal A}_1}},
\quad
K_{0,1}^\varsigma=-{\frac{\sqrt3\,(\L_{11}^\varsigma+\L_{21}^\varsigma)}{\sqrt2\,{\cal A}_2}},\\
&&K_{1,-1}^\varsigma={\frac{\sqrt3\, \L_{10}^\varsigma-\L_{20}^\varsigma-\sqrt2\, \L_{00}^\varsigma}
{\sqrt2\, {\cal A}_2}},
\quad
K_{1,0}^\varsigma={\frac{\sqrt3\,(\L_{21}^\varsigma-\L_{11}^\varsigma)}{\sqrt2\,{\cal A}_1}},
\quad
K_{1,1}^\varsigma=-{\frac{\sqrt3\,\L_{22}^\varsigma}{{\cal A}_2}}.
\end{eqnarray}

The coefficient $T^{\varsigma}_{\lambda\mu}(t)$ is connected with 
$\P^{\varsigma}_{\lambda\mu}(t)$ by the relation 
$\P^{\varsigma}_{\lambda\mu}(t)=AT^{\varsigma}_{\lambda\mu}(t)$, $A$ being 
the number of nucleons. 
\end{widetext}

\end{document}